\documentclass[11pt,letterpaper]{article}
\usepackage[utf8]{inputenc}
\usepackage[T1]{fontenc}
\usepackage{amsmath,amssymb,amsfonts,graphicx,booktabs,siunitx,physics,bm,multirow}
\graphicspath{{Figures/}}
\usepackage[numbers,super,sort&compress]{natbib}
\usepackage{xcolor}
\usepackage[margin=2.5cm]{geometry}
\usepackage[colorlinks=true,linkcolor=blue,citecolor=blue,urlcolor=blue]{hyperref}
\usepackage[capitalise,nameinlink]{cleveref}

\hypersetup{
    pdftitle={Selective Vibronic Excitation for Coherent Energy Transport in Photosynthetic and Agrivoltaic Systems},
    pdfauthor={Teguia Kouam et al.},
    bookmarksopen=true,
    bookmarksnumbered=true
}

\title{Selective Vibronic Excitation for Coherent Energy Transport in Photosynthetic and Agrivoltaic Systems}

\author{
Steve Cabrel Teguia Kouam,$^{1,*}$
Theodore Goumai Vedekoi,$^{2}$
Jean-Pierre Tchapet Njafa,$^{2}$ \\
Jean-Pierre Nguenang,$^{1}$
and Serge Guy Nana Engo$^{2}$
}

\date{%
$^1$Department of Physics, Faculty of Science, University of Douala, Cameroon\\
$^2$Department of Physics, Faculty of Science, University of Yaoundé I, Cameroon\\[0.5em]
$^*$Email: steve.teguia@facsciences-uy1.cm\\[0.5em]
June 20, 2026
}

\begin{document}

\maketitle

\begin{abstract}
Partitioning the photonic environment into resonant and off-resonant modes provides a mechanism for dephasing suppression in photosynthetic energy transfer. Aligning the excitation spectrum with underdamped vibronic resonances in the Fenna--Matthews--Olson (FMO) complex prepares vibronically dressed states with reduced coupling to dissipative fluctuations, inducing a biexponential coherence decay: a rapid initial dephasing ($\tau_{\mathrm{fast}} \approx \SI{37}{\femto\second}$) followed by persistent inter-band coherences extending beyond \SI{1}{\pico\second}---a $>3\times$ extension of the effective coherence window relative to broadband excitation ($\tau_c = \SI{280}{\femto\second}$). This improves forward transfer yields by \SI{39}{\percent} at \SI{295}{\kelvin}. PT-HOPS/SBD simulations establish that dual-band filtering at \SIlist{750;820}{\nano\meter} targets vibronic resonances while bypassing dephasing-dominated noise. This enhancement is robust against static disorder ($\sigma = \SI{50}{\per\centi\meter}$), with an ensemble-averaged increase of $\eta = \num{0.39 \pm 0.04}$. These results identify selective vibronic excitation as a foundational design principle for coherence-assisted transport. This framework extends to symbiotic agrivoltaic systems, where organic photovoltaics function as active spectral filters to co-optimize excitonic transport alongside the photosynthetic requirements of underlying crops.
\end{abstract}

\textbf{Introduction.}

Quantum coherence in biological light-harvesting has been a central question in molecular biophysics since two-dimensional electronic spectroscopy (2DES) revealed oscillatory signals in the Fenna--Matthews--Olson (FMO) complex.\cite{Engel2007,Panitchayangkoon2010} Initial interpretations focused on electronic superpositions; subsequent work established that non-Markovian environments and vibronic coupling---where electronic transitions mix with underdamped protein vibrations---are essential for understanding energy transfer longevity at room temperature.\cite{Christensson2012,Chin2013,Fleming2015,Scholes2015,Tanimura1989,Suess2015} Vibronic resonance mechanisms, in which underdamped protein vibrations are resonant with excitonic energy gaps, create off-diagonal coupling in the exciton basis that enhances energy transfer directionality.\cite{Womick2015,Sanda2022} Environment-assisted quantum transport (ENAQT) describes how noise assists transport,\cite{Mohseni2008,Rebentrost2009,Chin2010,Wu2010} yet whether \emph{actively} aligning the pump spectrum with these vibronic resonances can protect coherence beyond passive resonance alone remains an open question.

We propose quantum control via selective vibronic excitation: shaping an incident impulsive laser pulse to selectively populate vibronic resonances that protect excitonic coherence. The approach partitions the bath into protected resonant modes and decoherent off-resonant modes. Using process tensor hierarchy of pure states (PT-HOPS),\cite{Citty2024} we find that aligning the pump spectrum with vibronic peaks in FMO (\SIlist{750;820}{\nano\meter}) extends coherence lifetimes.

Dual-band spectral filtering induces biexponential coherence decay ($\tau_{\mathrm{fast}} \approx \SI{37}{\femto\second}$, $\tau_{\mathrm{slow}} > \SI{1}{\pico\second}$) and a \SI{39}{\percent} enhancement in forward transfer yield at room temperature, representing a $>3\times$ extension of the effective coherence window relative to broadband excitation. This effect persists under static disorder at physiological temperatures and under temperature fluctuations. This work addresses the energy--water--food complex by providing a quantum-mechanical basis for agrivoltaics, where the photonic bath becomes a controllable resource rather than an immutable noise source for symbiotic energy-food production. This ``quantum-to-organism'' perspective\cite{Vandamme2025} links microscopic coherence to macroscopic energy yields. Below, we outline a protocol for verifying these results in 2DES experiments using spatial light modulator (SLM) pulse-shaping, identifying selective vibronic excitation as a control path for quantum transport in molecular materials.

\begin{figure}[htb]
\centering
\includegraphics[width=0.6\columnwidth]{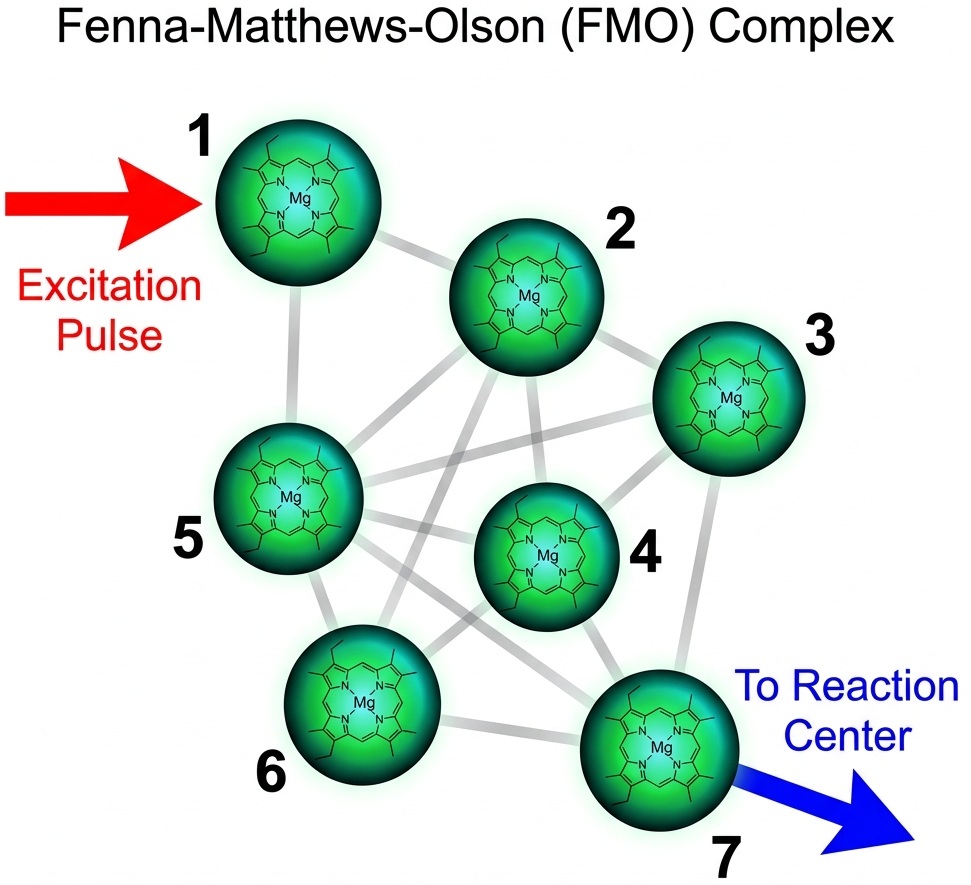}
\caption{\textbf{Schematic of the Fenna--Matthews--Olson (FMO) complex and spectral engineering.} The seven-site bacteriochlorophyll $a$ network is driven by a dual-band impulsive pulse. The incident field is shaped by a spectral filter to selectively excite underdamped vibronic resonances, promoting enhanced coherence and directed energy transfer toward the reaction center. Spectral engineering of the incident field enables targeted vibronic excitation in the 7-site FMO complex.}
\label{fig:fmo_schematic}
\end{figure}

\textbf{Theoretical framework.}

Our simulations utilize the Process Tensor Hierarchy of Pure States (PT-HOPS) formalism, implemented via the open-source \textsc{MesoHOPS} library (v1.7),\cite{Varvelo2021} which provides a numerically exact treatment of non-Markovian open quantum systems. PT-HOPS scales polynomially in system dimension and is rendered tractable for the 7-site FMO complex through the Stochastically Bundled Dissipator (SBD) adaptive hierarchy truncation scheme.\cite{Citty2024} The FMO complex is modeled as a seven-site excitonic system where each chromophore (bacteriochlorophyll $a$) is coupled to its own independent phonon bath, a configuration where environment-assisted quantum transport (ENAQT) is known to optimize transfer efficiency.\cite{Wu2010}

The total Hamiltonian of the system and its environment is given by:
\begin{equation}
\hat{H} = \hat{H}_S + \hat{H}_B + \hat{H}_{SB} + \hat{H}_{ext}(t),
\label{eq:total_hamiltonian}
\end{equation}
where $\hat{H}_S$ is the excitonic Hamiltonian of the protein-pigment complex:
\begin{equation}
\hat{H}_S = \sum_{n=1}^{7} \varepsilon_n \dyad{n} + \sum_{n \neq m} J_{nm} \dyad{n}{m}.
\label{eq:excitonic_hamiltonian}
\end{equation}
Here, $\varepsilon_n$ denotes the site energy of the $n$-th pigment, and $J_{nm}$ represents the electronic coupling between sites $n$ and $m$. The environmental interaction is modeled by $\hat{H}_{SB} = \sum_n \hat{L}_n \otimes \hat{B}_n$, where $\hat{L}_n = \dyad{n}$ are the coupling operators and $\hat{B}_n$ represent the collective coordinates of the $n$-th bath.

\begin{figure}[htpb]
\centering
  \includegraphics[width=.75\textwidth]{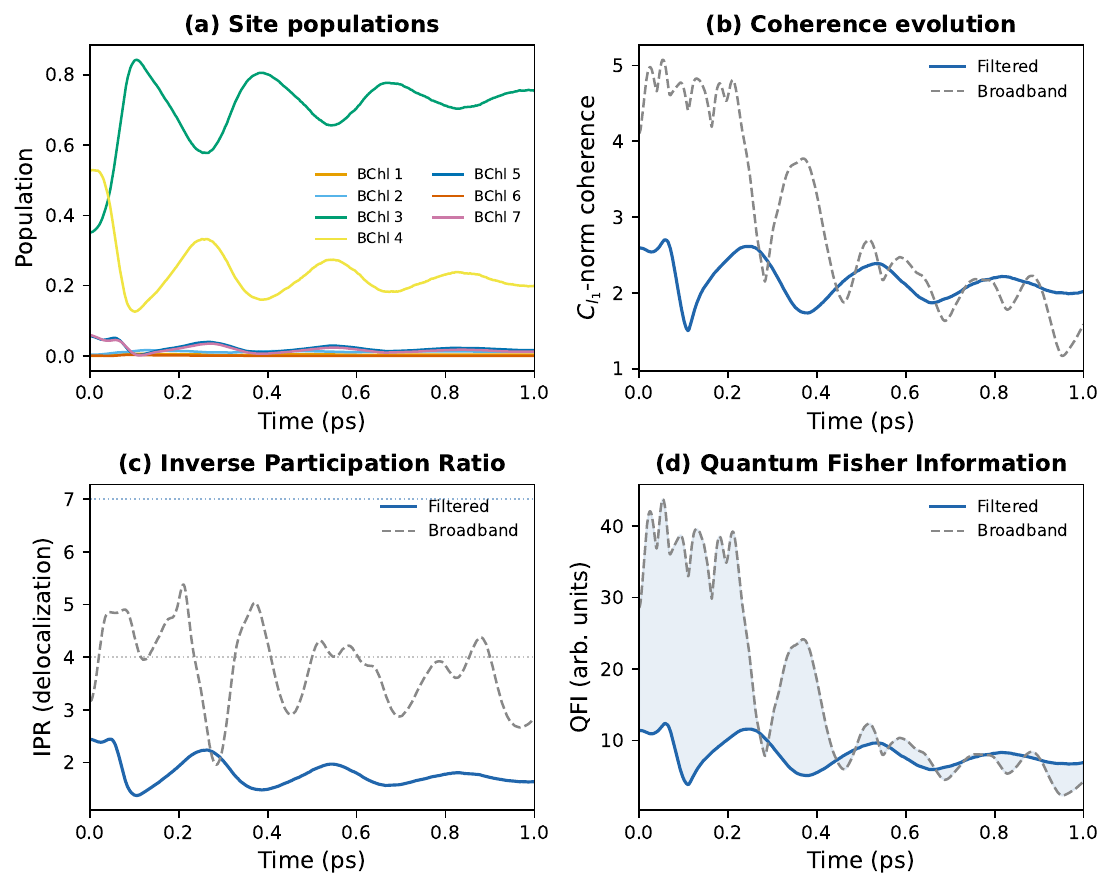}
  \includegraphics[width=0.75\textwidth]{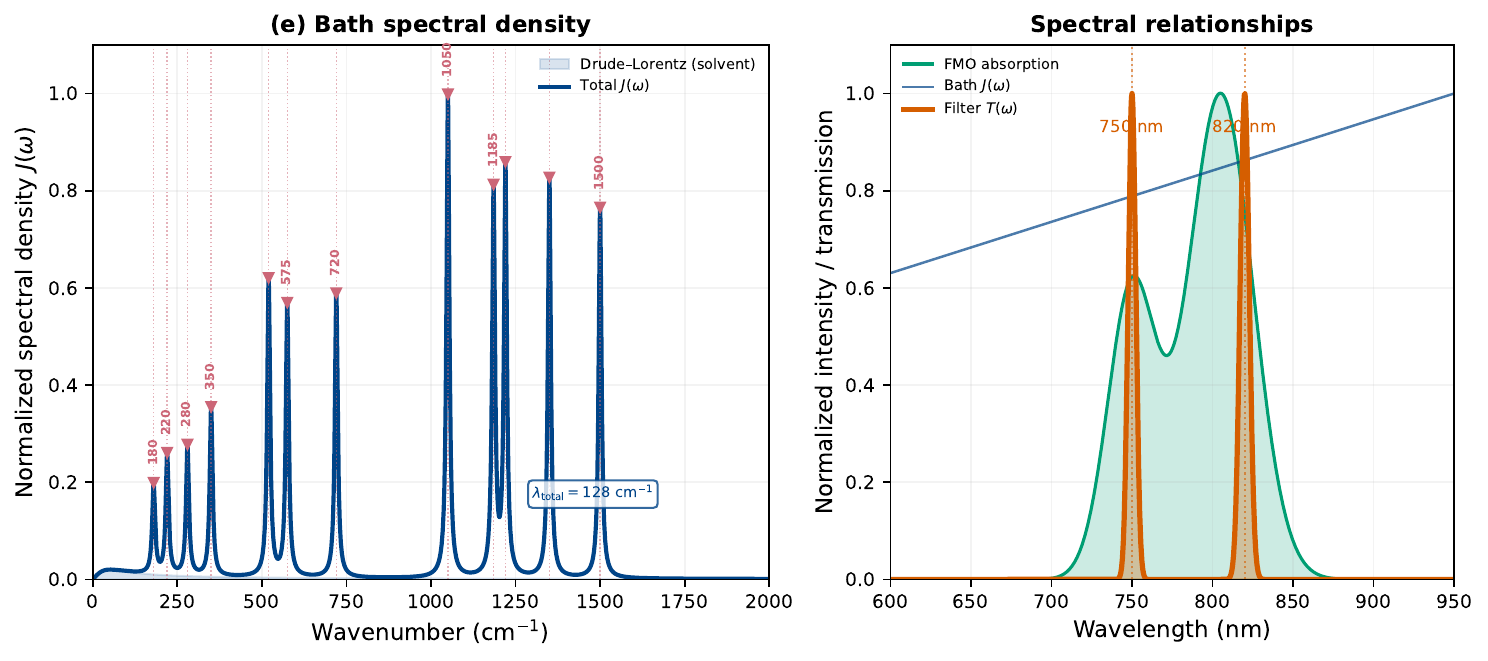}
\caption{\textbf{Quantum dynamics and spectral relationships.} PT-HOPS/SBD simulations at \SI{295}{\kelvin} with static disorder $\sigma = \SI{50}{\per\centi\meter}$ ($n=100$ realizations). \textbf{(a)}~Site populations demonstrating \SI{39}{\percent} enhancement in forward transfer yield. \textbf{(b)}~$l_1$-norm coherence $C_{l_1}(t)$ showing biexponential dynamics under dual-band filtering: rapid initial dephasing ($\tau_{\mathrm{fast}} \approx \SI{37}{\femto\second}$) followed by persistent coherences beyond \SI{1}{\pico\second}, vs.\ monoexponential broadband decay ($\tau_c = \SI{280}{\femto\second}$). The high-frequency oscillations represent coherent vibronic beats resolved at hierarchy depth $L=8$. Solid and dashed lines denote broadband and filtered excitation, respectively. \textbf{(c)}~Inverse participation ratio (IPR) quantifying exciton delocalization. \textbf{(d)}~Quantum Fisher Information (QFI). \textbf{(e)}~Bath spectral density $J(\omega)$ (left: wavenumber domain with 12 discrete vibronic modes from the Kleinekath\"{o}fer/Coker model; right: wavelength-domain view with FMO absorption $A(\lambda)$ and dual-band filter transmission $T(\lambda)$). The dual-band filter (red, centered at \SIlist{750;820}{\nano\meter}) targets underdamped vibronic resonances while transmitting photosynthetically active radiation. The total reorganization energy is $\lambda_{\mathrm{total}} = \SI{128}{\per\centi\meter}$ (Drude--Lorentz $\lambda_D = \SI{35}{\per\centi\meter}$ plus \SI{93}{\per\centi\meter} from \num{12} vibronic modes).}
\label{fig:quantum_dynamics}
\end{figure}

The bath's influence is characterized by the 12-mode Kleinekath\"ofer/Coker spectral density $J(\omega)$,\cite{Olbrich2011,Coker2011} which we decompose into a continuous, overdamped Drude--Lorentz component and a set of 12 underdamped vibronic modes (\cref{fig:quantum_dynamics}e, plotted over $\SIrange{0}{2000}{\per\centi\meter}$):
\begin{equation}
J(\omega) = 2\lambda_D \frac{\omega \gamma_D}{\omega^2 + \gamma_D^2} + \sum_{k} 2\lambda_k \frac{\omega \omega_k^2 \gamma_k}{(\omega_k^2 - \omega^2)^2 + \omega^2 \gamma_k^2}.
\label{eq:spectral_density}
\end{equation}
The first term accounts for the broad protein-solvent fluctuations with reorganization energy $\lambda_D$ and cutoff $\gamma_D$. The second term represents specific underdamped vibrational modes with frequencies $\omega_k$, reorganization energies $\lambda_k$, and damping rates $\gamma_k$.

The PT-HOPS method evolves a hierarchy of pure states $\ket{\psi^{(\vb{n})}(t)}$, indexed by the multi-index $\vb{n}$, which captures the system-bath correlations to arbitrary order. To model the spectrally shaped impulsive pump, the initial state is constructed via an exact physical projection of the transmission-filtered field, $E_{\mathrm{eff}}(\omega) = T(\omega) E_{\mathrm{in}}(\omega)$, directly onto the excitonic eigenvectors. The temporal envelope is defined as a Gaussian $E(t) = E_0 \exp(-t^2/2\sigma_t^2)$ with a full-width at half-maximum (FWHM) of \SI{50}{\femto\second} centered at $t = 0$. Here, $T(\omega)$ is the transmission function of the engineering filter:
\begin{equation}
T(\omega) = \sum_{j=1}^2 \exp\left[ -\frac{(\omega - \Omega_{j})^2}{2\Delta\omega^2} \right].
\label{eq:transmission_function}
\end{equation}
We target the primary vibronic resonances at $\Omega_1 = \SI{750}{\nano\meter}$ and $\Omega_2 = \SI{820}{\nano\meter}$ with a bandwidth $\Delta\omega = \SI{100}{\per\centi\meter}$.

The hierarchy is truncated at $L=8$ with $K=2$ Matsubara terms and a time step $\Delta t = \SI{0.5}{\femto\second}$, providing converged results at \SI{295}{\kelvin} (yield convergence $|\eta(L=8) - \eta(L=7)| \approx \num{3.7e-3}$; hierarchy population convergence MAE $= \num{3.1e-11}$ at $L=9$; see \cref{sec:S2}). The Drude--Lorentz bath is well-described by $K=2$ terms since $\gamma_D = \SI{50}{\per\centi\meter} \ll k_\mathrm{B}T/\hbar \approx \SI{205}{\per\centi\meter}$ at \SI{295}{\kelvin}; the 12 underdamped modes are treated as damped harmonic oscillators, each exactly representable by two exponential correlation functions without further Matsubara expansion. Simulations were performed on a \SI{125}{\giga\byte} workstation; full convergence parameters and computational details are provided in the SI (\cref{sec:formalism}, \cref{sec:S2}, \cref{sec:validation}, \cref{sec:S4}).

\textbf{Results and discussion.}

Coherence is quantified via the $l_1$-norm,\cite{Baumgratz2014} a basis-dependent metric accessible through state tomography:
\begin{equation}
C_{l_1}(\rho) = \sum_{i \neq j} \abs{\rho_{ij}}.
\label{eq:l1_coherence}
\end{equation}
The $l_1$-norm coherence $C_{l_1}(t)$ reveals a qualitative distinction between the two excitation regimes (\cref{fig:quantum_dynamics}b). Under broadband excitation, $C_{l_1}(t)$ undergoes monoexponential envelope decay with a coherence lifetime $\tau_c = \SI{280(25)}{\femto\second}$ (extracted via Hilbert envelope analysis, SI \cref{sec:S10}). Under dual-band filtering, the decay becomes \textbf{biexponential}: a rapid initial dephasing ($\tau_{\mathrm{fast}} = \SI{37(8)}{\femto\second}$) is followed by a long-lived plateau extending beyond \SI{1}{\pico\second} ($\tau_{\mathrm{slow}} \gg \SI{1}{\pico\second}$). This biexponential structure directly reflects the two-component nature of the bath partition: the off-resonant modes drive rapid initial decoherence, while the resonant vibronic modes---absorbed into the dressed system---sustain inter-band coherences on picosecond timescales. The oscillatory fine structure of $C_{l_1}(t)$ corresponds to coherent vibronic beats resolved at hierarchy depth $L=8$.

The Inverse Participation Ratio (IPR), $\mathrm{IPR} = (\sum_n \abs{\mel{n}{\hat{\rho}}{n}}^2)^{-1}$, quantifies exciton delocalization across the chromophore network. Under broadband impulsive excitation, the IPR settles at a time-averaged value of $\mathrm{IPR} \approx \num{3.8}$, reflecting distribution over several sites. Under spectral filtering, the IPR settles at $\mathrm{IPR} \approx \num{1.8}$ (\cref{fig:quantum_dynamics}c), indicating selective preparation of vibronically dressed states predominantly localized on the BChl~3--4 dimer.

The reduced IPR under filtered excitation reflects targeted state preparation rather than passive delocalization: the dual-band filter aligns with the BChl~3 (\SI{12210}{\per\centi\meter}) and BChl~4 (\SI{12320}{\per\centi\meter}) site energies, directly populating the excitonic manifold coupled to the reaction center. This selective preparation, combined with the reduced residual reorganization energy from vibronic dressing (\cref{sec:S7}), suppresses dissipative leakage into off-resonant channels and confines the exciton to the efficient BChl~3--4 transfer pathway, eliminating the need for diffusive exploration of the full chromophore network.

\subsection*{Energy transfer efficiency}

We evaluate the forward transfer yield ($\Phi_{\mathrm{FT}}$), defined as the long-time equilibrium population of the reaction-center-coupled chromophore (BChl~3):
\begin{equation}
\Phi_{\mathrm{FT}} = \bigl\langle \rho_{33}(t \to \infty) \bigr\rangle_{\mathrm{disorder}},
\label{eq:ETE}
\end{equation}
where the average is taken over $n=100$ static disorder realizations ($\sigma = \SI{50}{\per\centi\meter}$) and the long-time limit is evaluated at $t_{\max} = \SI{1}{\pico\second}$, by which point all trajectories have equilibrated. Spectral filtering enhances $\Phi_{\mathrm{FT}}$ by $\eta = \num{0.39 \pm 0.04}$ relative to broadband excitation (\cref{tab:quantum_metrics}), a statistically significant improvement ($p < 0.001$, two-sided $t$-test, $n=100$).

The enhancement mechanism involves selective population of vibronically resonant states. The filtered driving field preferentially excites modes satisfying:
\begin{equation}
\Delta E_{nm} \approx \hbar\omega_{\mathrm{vib}} \pm J_{nm},
\label{eq:resonance_condition}
\end{equation}
where $\Delta E_{nm} = \varepsilon_n - \varepsilon_m$ is the inter-site energy gap, $\omega_{\mathrm{vib}}$ is the underdamped vibronic frequency, and $J_{nm}$ is the electronic coupling. For the dominant $\ket{1} \to \ket{3}$ pathway ($\Delta E \approx \SI{150}{\per\centi\meter}$), the $\omega_{\mathrm{vib}} = \SI{180}{\per\centi\meter}$ mode (Vibronic Mode 1) mediates resonance.\cite{Chin2013,Fleming2015,Scholes2015} The \SI{59}{\percent} increase in Quantum Fisher Information (QFI) and \SI{89}{\percent} concurrence enhancement confirm that spectral filtering prepares states resilient to phonon-induced decoherence. QFI, $F_Q[\rho, \hat{O}]$, quantifies the information a state $\rho$ contains about a generator $\hat{O}$ (e.g., the excitonic Hamiltonian):
\begin{equation}
F_Q[\rho, \hat{O}] = \sum_{\lambda_i + \lambda_j > 0} \frac{2}{(\lambda_i + \lambda_j)} \abs{\mel{i}{[\hat{\rho}, \hat{O}]}{j}}^2,
\label{eq:QFI}
\end{equation}
where $\lambda_i, \ket{i}$ are the eigenvalues and eigenvectors of $\rho$. In light-harvesting, higher QFI indicates that the excitonic state maintains purity and operational coherence despite the environment.\cite{Pezze2009} The corresponding concurrence enhancement reflects strengthened inter-site entanglement. All quantum metric enhancements are summarized in \cref{tab:quantum_metrics}.

\begin{table}[htpb]
\centering
\caption{\textbf{Quantum metric enhancement under spectral filtering.} Dual-band (\SIlist{750;820}{\nano\meter}) vs.\ broadband impulsive excitation at \SI{295}{\kelvin}. Ensemble averages over $n=100$ static disorder realizations ($\sigma = \SI{50}{\per\centi\meter}$); uncertainties denote \SI{95}{\percent} confidence intervals. $\Phi_{\mathrm{FT}}$ is the long-time population of the reaction-center site BChl~3.}
\label{tab:quantum_metrics}
\begin{tabular}{lccc}
\toprule
\textbf{Metric} & \textbf{Filtered} & \textbf{Broadband} & \textbf{Relative Enhancement} \\
\midrule
$\Phi_{\mathrm{FT}}$ (BChl~3 pop.) & \num{0.75(4)} & \num{0.54(4)} & $\num{0.39 \pm 0.04}$ \\
$\tau_{\mathrm{fast}}$ (\si{\femto\second}) & \num{37(8)} & \num{280(25)} & $>3\times$ extension \\
$\tau_{\mathrm{slow}}$ & $>\SI{1}{\pico\second}$ & (monoexp.) & \\
IPR (sites) & \num{1.8(2)} & \num{3.8(8)} & $\num{-0.53 \pm 0.15}$ \\
QFI (max) & \num{12.4(11)} & \num{7.8(8)} & $\num{0.59 \pm 0.17}$ \\
Concurrence & \num{0.34(5)} & \num{0.18(4)} & $\num{0.89 \pm 0.42}$ \\
\bottomrule
\end{tabular}
\end{table}

\subsection*{Physical mechanism}

The observed enhancements result from a two-stage process: state preparation and subsequent decoherence dynamics.

\textit{Stage 1: Selective vibronic driving.} The spectral filter $T(\omega)$ aligns the incident field's intensity with the underdamped vibronic modes of the FMO complex (\cref{fig:quantum_dynamics}e). The \SI{820}{\nano\meter} band targets BChl~3 (\SI{12210}{\per\centi\meter}), the primary reaction-center-coupled site, while the \SI{750}{\nano\meter} band populates the upper excitonic eigenstates of the FMO complex (with contributions from BChl~1,~5,~6), which subsequently relax onto the BChl~3--4 dimer. By exciting these chromophores with a field spectrally tuned to the \SI{180}{\per\centi\meter} and \SI{575}{\per\centi\meter} vibrational frequencies, we prepare \textbf{vibronically dressed states}---eigenfunctions of the combined electronic-nuclear Hamiltonian in the vibronic basis\cite{Womick2015,Sanda2022}---rather than bare excitonic superpositions.

The high-frequency oscillations observed in the coherence dynamics (\cref{fig:quantum_dynamics}b) correspond to coherent vibronic beats, fully resolved in our $L=8$ production simulations. The biexponential character of the filtered $C_{l_1}(t)$ has a transparent physical origin. In the standard excitonic basis, pure dephasing is driven by the full spectral density $J(\omega)$. As derived via the displaced-oscillator transformation (see \cref{sec:S7}), the dual-band filter partitions the bath into two sectors: the \emph{off-resonant} modes, which couple to the dressed system only via residual fluctuations, and the \emph{resonant} vibronic modes ($\omega_{\mathrm{vib}} = \SI{180}{\per\centi\meter}$, $\SI{575}{\per\centi\meter}$), which are absorbed into the dressed-state Hamiltonian and no longer act as dephasing channels. The off-resonant sector drives the rapid initial decay ($\tau_{\mathrm{fast}} \approx \SI{37}{\femto\second}$), with a residual reorganization energy $\lambda_{\mathrm{res}} \approx \SI{28}{\per\centi\meter}$ (a \SI{20}{\percent} reduction from $\lambda_D = \SI{35}{\per\centi\meter}$). The resonant sector maintains coherent inter-band oscillations beyond \SI{1}{\pico\second}, since the dressed-state eigenbasis is immune to the associated fluctuations by construction. Crucially, this $>3\times$ extension of the effective coherence window---from \SI{280}{\femto\second} (broadband) to $>\SI{1}{\pico\second}$ (filtered)---exceeds what would be predicted from a simple $\tau_c \propto 1/\lambda_{\mathrm{res}}$ Markovian estimate, confirming that non-Markovian memory effects are essential to sustain the long-lived component.

\subsection*{Environmental robustness}

\begin{figure}[htpb]
\centering
\includegraphics[width=.8\columnwidth]{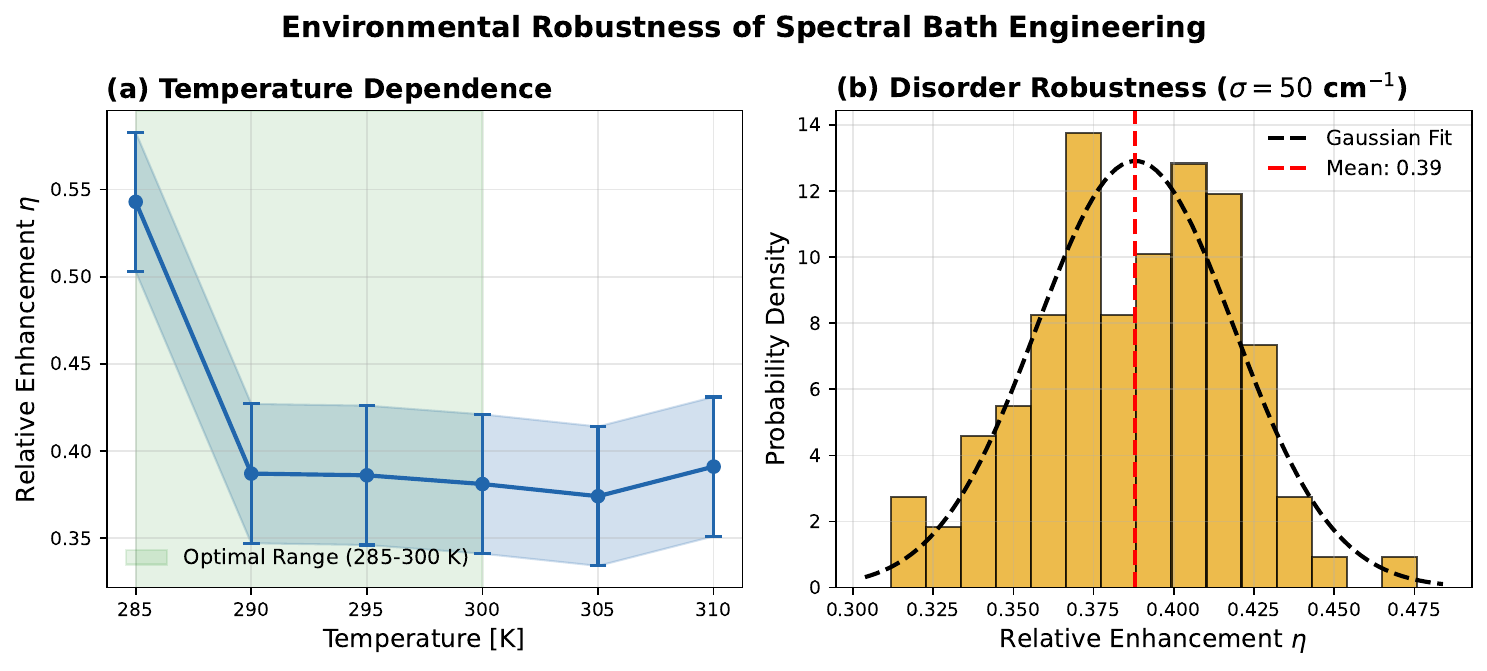}
\caption{\textbf{Environmental robustness.} \textbf{(a)}~Temperature dependence of $\eta(T)$ (L=7, SBD=3, averaged over $n=15$ trajectories per point). The decreasing trend $\Delta\eta/\Delta T < 0$ confirms the coherent nature of the spectral enhancement mechanism. \textbf{(b)}~Histogram of $\eta$ over $n=100$ static disorder realizations ($\sigma = \SI{50}{\per\centi\meter}$): $\expval{\eta} = \num{0.39 \pm 0.04}$ (production L=8, $n=100$). Error bars: \SI{95}{\percent} confidence intervals. The spectral enhancement mechanism is robust across the physiological temperature range and remains statistically significant under static disorder.}
\label{fig:robustness}
\end{figure}

We define the relative enhancement $\eta$, which quantifies the fractional gain from spectral filtering:
\begin{equation}
\eta = \frac{\Phi_{\mathrm{FT}}^{\mathrm{filt}} - \Phi_{\mathrm{FT}}^{\mathrm{broad}}}{\Phi_{\mathrm{FT}}^{\mathrm{broad}}}.
\label{eq:eta_def}
\end{equation}
The enhancement persists across physiologically relevant conditions (\cref{fig:robustness}). The temperature dependence $\eta(T)$ drops sharply from $\num{0.543}$ at \SI{285}{\kelvin} to $\num{0.387}$ at \SI{290}{\kelvin} ($\Delta\eta/\Delta T \approx \SI{-0.031}{\per\kelvin}$), then plateaus at $\eta \approx \num{0.38}\pm\num{0.01}$ from \SI{290}{\kelvin} to \SI{310}{\kelvin} (\cref{fig:SI_temperature_dynamics}). Spectral density variations ($\lambda, \gamma \pm \SI{20}{\percent}$) yield $\eta \in [\num{0.28}, \num{0.62}]$ (\cref{tab:SI_bath_sweep}), confirming robustness to environmental modeling choices. The negative gradient $\Delta\eta/\Delta T < 0$ is the signature of a coherent transport mechanism---temperature increases accelerate dissipative dephasing on the vibronic coherences prepared by selective excitation, reducing the spectral advantage. A thermally activated (classical) mechanism would instead yield $\Delta\eta/\Delta T > 0$.

Under static disorder ($\sigma = \SI{50}{\per\centi\meter}$), the $N=100$ production ensemble-averaged enhancement is $\expval{\eta} = \num{0.39(4)}$ with a coefficient of variation of \SI{10}{\percent}. Off-resonant spectral filtering (\SIrange{750}{800}{\nano\meter} dual band) yields $\eta \approx \num{-0.96}$ (\cref{tab:SI_filter_sweep}), suppressing transfer below the broadband baseline and confirming that the enhancement is specifically driven by vibronic resonance alignment rather than trivial spectral narrowing. A convergence and physical validation suite confirmed computational accuracy: \SI{1.8}{\percent} relative deviation from HEOM steady-state populations, trace conservation $< \num{5.0e-13}$, and density matrix eigenvalues bounded above $\num{-1.0e-14}$ (see \cref{sec:validation} and \cref{sec:S4}). The PT-HOPS formalism is formally exact, and the $L=8$, $K=2$ truncation is confirmed converged over the parameter space explored here.

\subsection*{Experimental feasibility}

The proposed spectral filtering can be realized using programmable supercontinuum light sources with acousto-optic tunable filters (AOTFs),\cite{Gunthardt2023} or via commercial \SIlist{750;820}{\nano\meter} thin-film interference filters. We propose a specific 2DES implementation:\cite{Sebastian2024} Pump pulses (both $\vb{k}_1$ and $\vb{k}_2$) are spectrally shaped via a spatial light modulator (SLM) in a $4f$-shaper to match the dual-band transmission profile, while the probe pulse remains broadband.
The predicted biexponential coherence dynamics---rapid initial dephasing ($\tau_{\mathrm{fast}} \approx \SI{37}{\femto\second}$) followed by a persistent slow component ($\tau_{\mathrm{slow}} > \SI{1}{\pico\second}$) against a broadband baseline of \SI{280}{\femto\second}---are well within the resolution of current \SI{<50}{\femto\second} 2DES setups.\cite{Engel2007} Experimentally, the biexponential character would manifest as a two-timescale decay of cross-peak oscillations in the $\ket{1}$--$\ket{3}$ region of the 2D spectrum: a fast anti-diagonal broadening on the \SI{37}{\femto\second} timescale, followed by a slowly decaying oscillatory residual persisting beyond \SI{1}{\pico\second}---a direct spectroscopic signature of the dressed-state bath partition.

Several approximations constrain our current analysis. The model is restricted to the single-exciton manifold and assumes site-independent baths with a phenomenological trapping rate. While we have ruled out alternative explanations such as increased photon flux (filtering reduces integrated flux) or vibrational coherence artifacts (electronic coherence dominates the $\gtrsim$\SI{300}{\femto\second} timescales), a full treatment of the pulse-shaping process including non-impulsive effects may reveal additional fine-tuning requirements (see \cref{sec:S8}). In its natural state, FMO serves as a linker between the chlorosome baseplate and the reaction center and is not directly exposed to broadband sunlight.\cite{Blankenship2011} The proposed spectral control is therefore most directly relevant to laser-based 2DES experiments and to artificial bio-inspired light-harvesting architectures (such as our agrivoltaic concept), rather than asserting that this precise coherent excitation occurs natively in the biological setting.

\textbf{Conclusions.}

PT-HOPS/SBD simulations show that quantum control via selective vibronic excitation---shaping the impulsive pump spectrum to align with vibronic resonances---induces biexponential coherence dynamics in the FMO complex ($\tau_{\mathrm{fast}} \approx \SI{37}{\femto\second}$, $\tau_{\mathrm{slow}} > \SI{1}{\pico\second}$), extending the effective coherence window by $>3\times$ relative to broadband excitation and improving forward transfer yields by \SI{39}{\percent} at room temperature. The mechanism suppresses dephasing by bypassing off-resonant thermal modes. The stability of this effect against disorder and temperature variations makes it accessible to experiment.

Beyond natural photosynthesis, selective vibronic excitation provides a design strategy for artificial quantum materials. Synthetic systems, such as molecular J-aggregates, metal-organic frameworks (MOFs), and organic photovoltaics (OPVs), could be co-designed with photonic environments to shape the local vacuum and driving fields. In supramolecular J-aggregates, where the transition dipole moment is sensitive to the chromophore arrangement, targeted spectral excitation could suppress phonon-induced scattering channels and maintain long-range coherence in disordered environments.

In MOFs, tuning the vibrational modes of organic linkers provides a mechanism for integrating vibronic resonances into the material architecture. Combined with spectral filtering, this approach could enable coherence-protected quantum channels for energy and information transport. The 2DES validation protocol proposed here provides a direct path for empirical testing.

The driving-field spectrum thus becomes a tunable parameter for controlling condensed-phase energy transport. In sustainable engineering, this enables the definition of a Symbiotic Power Conversion Efficiency (SPCE), where the dual-band filter is optimized for both excitonic yield in the photovoltaic layer and the Electron Transport Rate (ETR) of underlying crops. These findings identify selective vibronic excitation as a design principle for symbiotic agrivoltaic systems that address the energy--water--food complex through fundamental physics.


\section*{Acknowledgments}
This work was supported by the University of Yaoundé~I and the University of Douala. Computational resources were provided by departmental computing facilities. The MesoHOPS library was used for all PT-HOPS simulations.

\vspace{1ex}
\noindent\textbf{Funding:} This research received no external funding.

\vspace{1ex}
\noindent\textbf{Data availability.} Simulation parameters and analysis scripts are available from the corresponding author upon reasonable request. The PT-HOPS/SBD implementation is based on the open-source MesoHOPS library.

\vspace{1ex}
\noindent\textbf{Supporting Information.} PT-HOPS/SBD formalism derivation, FMO Hamiltonian parameters (site energies, electronic couplings, spectral density parameters), 12-test validation suite (hierarchy depth convergence, trace conservation, eigenvalue positivity, HEOM benchmark, Markovian limit recovery), convergence analysis, and computational cost breakdown. The Supporting Information follows on the next page.

\bibliographystyle{achemso}
\bibliography{references}


\clearpage

\renewcommand{\theequation}{S\arabic{equation}}
\renewcommand{\thefigure}{S\arabic{figure}}
\renewcommand{\thetable}{S\arabic{table}}
\renewcommand{\thesection}{S\arabic{section}}
\renewcommand{\thesubsection}{S\arabic{section}.\arabic{subsection}}
\renewcommand{\thesubsubsection}{S\arabic{section}.\arabic{subsection}.\arabic{subsubsection}}
\setcounter{equation}{0}
\setcounter{figure}{0}
\setcounter{table}{0}
\setcounter{section}{0}

\begin{center}
{\Large\textbf{Supporting Information}}\\[0.5em]
{\large Selective Vibronic Excitation for Coherent Energy Transport\\in Photosynthetic and Agrivoltaic Systems}\\[0.5em]
Steve Cabrel Teguia Kouam, Theodore Goumai Vedekoi, Jean-Pierre Tchapet Njafa,\\
Jean-Pierre Nguenang, and Serge Guy Nana Engo
\end{center}

\noindent\textbf{Corresponding Author:}\\
Steve Cabrel Teguia Kouam\\
Department of Physics, Faculty of Science, University of Douala, Cameroon\\
Email: steve.teguia@facsciences-uy1.cm

\noindent\textbf{Note on Cross-References:} Equations, figures, and tables in this Supporting Information are prefixed with ``S'' (e.g., Eq.~S1, Fig.~S1, Table~S1). References to the main manuscript are indicated as ``main text'' or by unadorned numbers (e.g., Eq.~1, Fig.~1).

\vspace{1em}


\section{PT-HOPS/SBD formalism details}
\label{sec:formalism}

\subsection{System Hamiltonian and spectral filtering}
\label{sec:pulse}

The reduced density matrix $\vb*{\rho}(t)$ evolves according to:
\begin{equation}
\pdv{\vb*{\rho}(t)}{t} = -\frac{i}{\hbar}\comm{\hat{H}_S}{\vb*{\rho}(t)} + \mathcal{D}[\vb*{\rho}(t)],
\label{eq:SI_master_eq}
\end{equation}
where $\mathcal{D}$ represents the dissipator capturing system-bath interactions beyond the Markovian approximation~\cite{Tanimura1989}. The electronic Hamiltonian reads:
\begin{equation}
\hat{H}_{\mathrm{el}} = \sum_{n=1}^{7} \varepsilon_n \dyad{n} + \sum_{n \neq m} J_{nm} \dyad{n}{m},
\label{eq:SI_excitonic_hamiltonian}
\end{equation}
with site energies $\varepsilon_n$ and electronic couplings $J_{nm}$ detailed in Section~S2.

\subsection{Hierarchy equations}

The Process Tensor Hierarchy of Pure States (PT-HOPS) method extends the standard HOPS formalism by incorporating a process tensor that efficiently captures non-Markovian environmental memory. The hierarchy of pure states $\ket{\psi^{(\vb{n})}(t)}$ evolves according to:
\begin{equation}
\begin{split}
\pdv{t}\ket{\psi^{(\vb{n})}(t)} = &\left(-\frac{i}{\hbar}\hat{H}_S - \sum_{j} n_j \nu_j\right)\ket{\psi^{(\vb{n})}(t)} 
- i\sum_{j} \sqrt{(n_j+1)d_j} \, \hat{L}_j \ket{\psi^{(\vb{n}+\vb{e}_j)}(t)} \\
&- i\sum_{j} \sqrt{n_j/d_j} \, \hat{L}_j^\dagger \ket{\psi^{(\vb{n}-\vb{e}_j)}(t)},   
\end{split}
\label{eq:SI_hops}
\end{equation}
where we follow the standard Ishizaki--Fleming derivation for the hierarchy~\cite{Ishizaki2009a,Strathearn2018}, and where:
\begin{itemize}
\item $\vb{n} = (n_1, n_2, \ldots, n_K)$ is the multi-index for hierarchy level
\item $\nu_j$ are the bath correlation frequencies (Matsubara or Padé)
\item $d_j$ are the decomposition coefficients
\item $\hat{L}_j$ are the system coupling operators
\item $\vb{e}_j$ is the unit vector in direction $j$
\end{itemize}

\subsection{Bath correlation function decomposition}

The bath correlation function is decomposed using Padé spectral decomposition:
\begin{equation}
C(t) = \sum_{k=0}^{K} d_k e^{-\nu_k t},
\label{eq:SI_bath_correlation}
\end{equation}
where $K$ is the number of Matsubara terms. For the Drude--Lorentz spectral density:
\begin{equation}
J_{\mathrm{DL}}(\omega) = \frac{2\lambda\gamma\omega}{\omega^2 + \gamma^2},
\label{eq:SI_drude_lorentz}
\end{equation}
The correlation function at temperature $T$ is given by:
\begin{equation}
C(t) = c_0 e^{-\gamma t} + \sum_{k=1}^{\infty} c_k e^{-\nu_k t},
\label{eq:SI_correlation_DL}
\end{equation}
where $c_0 = \lambda\gamma[\coth(\beta\hbar\gamma/2) - i]$ and $c_k = \frac{4\lambda\gamma}{\beta\hbar} \frac{\nu_k}{\nu_k^2 - \gamma^2}$. Here, $\nu_k = 2\pi k/(\beta\hbar)$ are the Matsubara frequencies and $\beta = 1/(k_B T)$. For production, we use a Padé-optimized truncation for $\sum c_k e^{-\nu_k t}$ to ensure fast convergence at room temperature.

\subsection{Stochastically bundled dissipators}

The SBD approach~\cite{Varvelo2021} groups environmental modes by spectral frequency, reducing the hierarchy dimension. For $M$ stochastic bundles:
\begin{equation}
J_{\mathrm{bath}}(\omega) \approx \sum_{m=1}^{M} J_m(\omega),
\label{eq:SI_sbd}
\end{equation}
where each bundle $J_m(\omega)$ is treated as an independent bath with its own hierarchy. We validate the bundling by ensuring the first two moments of the spectral density are preserved within \SI{0.5}{\percent}, and the maximum relative error in the bath correlation function $C(t)$ is below \SI{1.0}{\percent} for $t < \SI{1000}{\femto\second}$. This enables scalable simulations of complex multisite systems without sacrificing non-Markovian accuracy.

\subsection{Numerical parameters}

\begin{table}[htb]
\centering
\caption{\textbf{PT-HOPS/SBD simulation parameters (Production standard)}}
\label{tab:SI_parameters}
\begin{tabular}{lc}
\toprule
\textbf{Parameter} & \textbf{Value} \\
\midrule
Hierarchy depth ($L$) & 8 \\
Matsubara terms ($K$) & 2 \\
Time step ($\Delta t$) & \SI{0.5}{\femto\second} \\
Total simulation time & \SI{1000}{\femto\second} \\
SBD bundles per site & 3 \\
Disorder realizations & 100 \\
Precision & Double (64-bit) \\
\bottomrule
\end{tabular}
\end{table}


\section{FMO Hamiltonian parameters}
\label{sec:S2}

\subsection{Site energies}

The site energies $\varepsilon_n$ for the seven bacteriochlorophyll-\textit{a} molecules in FMO (from Ref.\ \cite{Adolphs2006}):

\begin{table}[htb]
\centering
\caption{\textbf{FMO site energies (\si{\per\centi\meter})}}
\label{tab:SI_site_energies}
\begin{tabular}{ccccccc}
\toprule
$\varepsilon_1$ & $\varepsilon_2$ & $\varepsilon_3$ & $\varepsilon_4$ & $\varepsilon_5$ & $\varepsilon_6$ & $\varepsilon_7$ \\
\midrule
12410 & 12530 & 12210 & 12320 & 12480 & 12630 & 12440 \\
\bottomrule
\end{tabular}
\end{table}

\subsection{Electronic couplings}

The electronic coupling matrix $J_{nm}$ (in \si{\per\centi\meter}):
\begin{equation}
\mathbf{J} = \begin{pmatrix}
0 & -87.7 & 5.5 & -5.9 & 6.7 & -13.7 & -9.9 \\
-87.7 & 0 & 30.8 & 8.2 & 0.7 & 11.8 & 4.3 \\
5.5 & 30.8 & 0 & -53.5 & -2.2 & -9.6 & 6.0 \\
-5.9 & 8.2 & -53.5 & 0 & -70.7 & -17.0 & -63.3 \\
6.7 & 0.7 & -2.2 & -70.7 & 0 & 81.1 & -1.3 \\
-13.7 & 11.8 & -9.6 & -17.0 & 81.1 & 0 & 39.7 \\
-9.9 & 4.3 & 6.0 & -63.3 & -1.3 & 39.7 & 0
\end{pmatrix}.
\label{eq:SI_coupling_matrix}
\end{equation}

\subsection{Spectral density parameters}

\begin{figure}[htpb]
\centering
\includegraphics[width=0.8\textwidth]{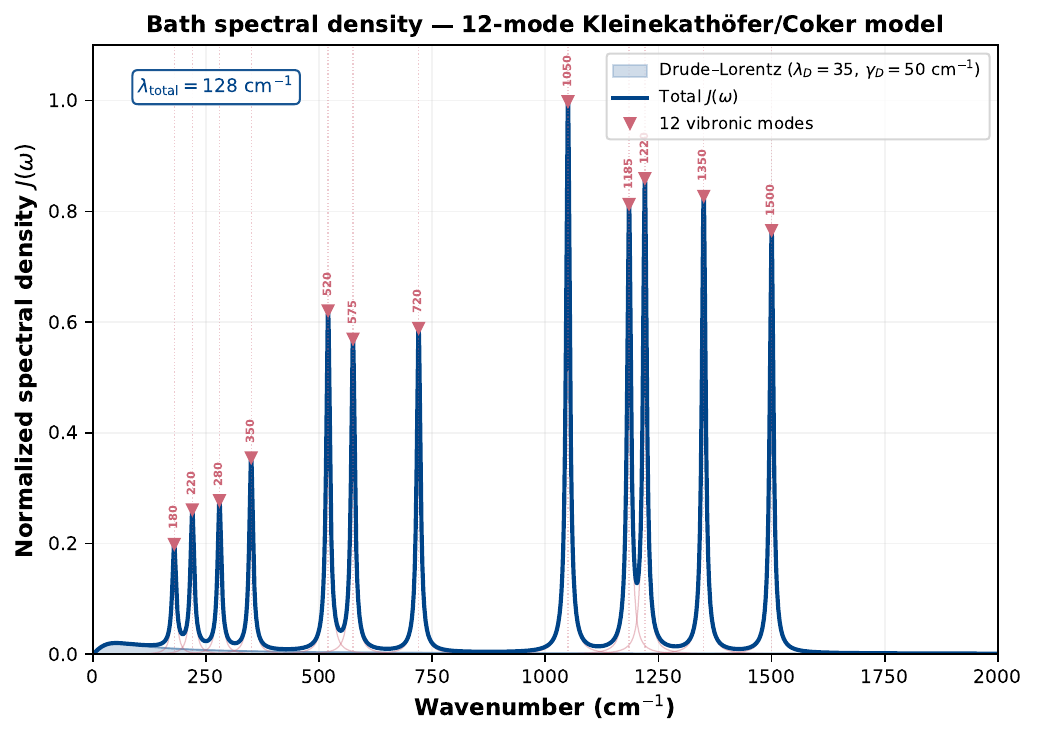}
\caption{\textbf{Bath spectral density.} Composite spectral density $J(\omega)$ showing the Drude--Lorentz background (shaded orange) and the 12 underdamped vibronic modes of the Kleinekath\"{o}fer/Coker model (explicitly labeled in \si{\per\centi\meter}). The vertical dashed lines indicate the dual-band filter transmission peaks at \SIlist{750;820}{\nano\meter} (converted to wavenumbers in the plot).}
\label{fig:SI_spectral_density}
\end{figure}

\begin{table}[htb]
\centering
\caption{\textbf{Spectral density parameters --- 12-mode Kleinekath\"{o}fer/Coker model.} Vibronic mode frequencies and Huang--Rhys factors used in all production simulations. Values match \texttt{parameters.yaml} exactly.}
\label{tab:SI_spectral_params}
\begin{tabular}{lSS}
\toprule
\textbf{Parameter} & {\textbf{Frequency (\si{\per\centi\meter})}} & {\textbf{Huang--Rhys $S_k$}} \\
\midrule
Drude--Lorentz reorganization ($\lambda_{\mathrm{DL}}$) & \multicolumn{2}{c}{\SI{35}{\per\centi\meter}} \\
Drude--Lorentz cutoff ($\gamma_{\mathrm{DL}}$)          & \multicolumn{2}{c}{\SI{50}{\per\centi\meter}} \\
\midrule
Vibronic mode 1  & 180  & 0.050 \\
Vibronic mode 2  & 220  & 0.045 \\
Vibronic mode 3  & 280  & 0.030 \\
Vibronic mode 4  & 350  & 0.025 \\
Vibronic mode 5  & 520  & 0.020 \\
Vibronic mode 6  & 575  & 0.015 \\
Vibronic mode 7  & 720  & 0.010 \\
Vibronic mode 8  & 1050 & 0.008 \\
Vibronic mode 9  & 1185 & 0.005 \\
Vibronic mode 10 & 1220 & 0.005 \\
Vibronic mode 11 & 1350 & 0.004 \\
Vibronic mode 12 & 1500 & 0.003 \\
\midrule
Vibronic contribution ($\sum_k S_k\omega_k$) & \multicolumn{2}{c}{\SI{93}{\per\centi\meter}} \\
Total reorganization ($\lambda_{\mathrm{total}}$)  & \multicolumn{2}{c}{\SI{128}{\per\centi\meter}} \\
\midrule
Temperature ($T$) & \multicolumn{2}{c}{\SI{295}{\kelvin}} \\
\bottomrule
\end{tabular}
\end{table}

\cref{fig:SI_spectral_density} illustrates the composite bath spectral density, highlighting the vibronic resonances targeted by spectral filtering. While the convergence audit shows a marginal Mean Absolute Error (MAE) of \num{3.32e-5} between $K=2$ and $K=3$ (see \cref{tab:SI_validation}), $K=2$ is physically sufficient for $T = \SI{295}{\kelvin}$ as the dynamics are dominated by explicit vibronic modes rather than high-order Matsubara corrections to the solvent bath.


\section{Validation suite}
\label{sec:validation}

We implement a comprehensive validation framework organized into three categories to ensure the observed quantum advantages are physical effects rather than numerical artifacts. Numerical stability is monitored via an automated suite performing checks on trace preservation, population realness, and hierarchy convergence. Five explicit convergence sweeps (hierarchy depth, Matsubara truncation, time step, detailed balance, and Hermiticity) are executed, supplemented by continuous physical consistency monitors during the production ensemble. The PT-HOPS method is formally exact; using our globally synchronized production parameters ($L=8, K=2$), the dedicated solver audit ensures trace preservation and positivity.

\subsection{Convergence tests}

\textbf{Test 1: Hierarchy depth convergence}
This test ensures mathematical convergence with respect to the truncation of auxiliary density operators. Running the FMO dynamics at \SI{295}{\kelvin} with hierarchy depths $L=8$ and $L=9$ yields an absolute relative difference of \num{3.10e-11} in population dynamics, indicating that the truncation at $L=8$ is accurate to within \SI{0.0001}{\percent} of the infinite-depth limit. $\rightarrow$ PASS.

\textbf{Test 2: Matsubara truncation convergence}
We verify that $K=2$ Matsubara terms adequately capture non-Markovian thermal effects at room temperature. Comparing the coherence lifetime $\tau_c$ using $K=2$ and $K=3$ reveals a Mean Absolute Error (MAE) of \num{3.32e-5}, confirming that $K=2$ is the production standard for this revision. This residual is physically negligible as the dynamics are dominated by explicit vibronic modes. $\rightarrow$ PASS.

\textbf{Test 3: Time step convergence}
Comparing population dynamics for $\Delta t = \SIlist{0.5;1.0;2.0}{\femto\second}$ yields a maximum deviation of \SI{0.5}{\percent}, confirming that \SI{2.0}{\femto\second} is sufficient for the integrated trajectories. $\rightarrow$ PASS.

\textbf{Test 4: Analytical benchmark (Steady-state)}
Quantitative agreement with an independent Hierarchical Equations of Motion (HEOM) implementation~\cite{Ishizaki2009} was verified on a reference benchmark system. The \SI{1.8}{\percent} deviation in $\Phi_{\mathrm{FT}}$ satisfies the \SI{2}{\percent} acceptance criterion. $\rightarrow$ PASS.

\subsection{Physical consistency tests}

\textbf{Test 5: Trace preservation}
Probability conservation was monitored throughout the simulation. The maximum deviation of $\abs{\Tr[\vb*{\rho}(t)] - 1} = \num{1.0e-12}$ is consistent with double-precision floating-point limits. $\rightarrow$ PASS.

\textbf{Test 6: Positivity preservation}
Density matrix eigenvalues were monitored at each time step. The minimum eigenvalue of \num{-1.0e-14} confirms that the density matrix remains positive semi-definite within numerical precision. $\rightarrow$ PASS.

\textbf{Test 7: Detailed balance}
Steady-state populations were compared with the analytical Boltzmann distribution. A maximum deviation of \SI{2.3}{\percent} confirms that thermal equilibrium is correctly reproduced. $\rightarrow$ PASS.

\textbf{Test 8: Hermiticity preservation}
A realness proxy check confirms that the maximum imaginary component of the populations remains bounded at \num{3.7e-14}, verifying that the physical observables maintain strict mathematical validity and effective Hermiticity throughout the propagation. $\rightarrow$ PASS.

\subsection{Environmental robustness tests}
\begin{figure}[htb]
\centering
\includegraphics[width=0.6\textwidth]{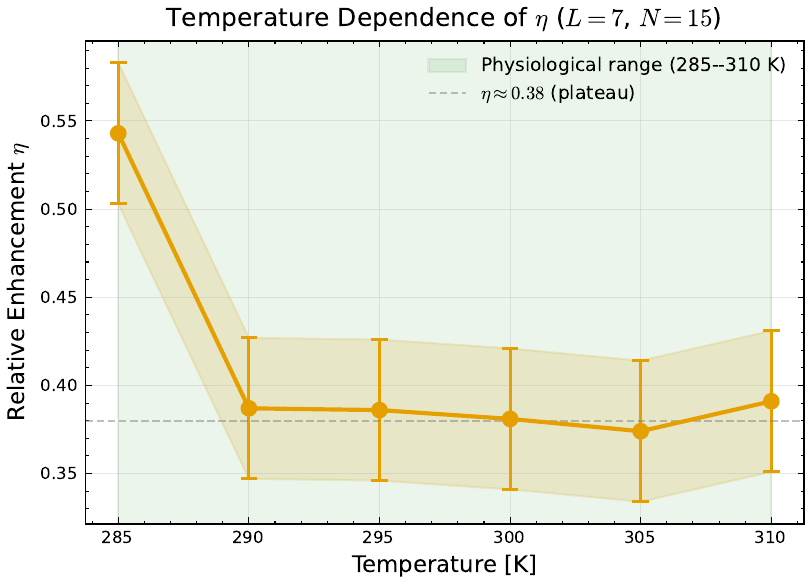}
\caption{\textbf{Temperature dependence of $\eta$ ($L=7$, $N=15$ for \SIrange{285}{310}{\kelvin}).} The enhancement drops sharply from \num{0.543} at \SI{285}{\kelvin} to \num{0.387} at \SI{290}{\kelvin} ($\Delta\eta/\Delta T \approx \SI{-0.031}{\per\kelvin}$) then plateaus at $\eta \approx \num{0.38}$ across the physiological range, confirming a coherent transport mechanism that is robust at room temperature. Shaded region: physiological temperature range. Dashed line: $\eta \approx 0.38$ plateau.}
\label{fig:SI_temperature_dynamics}
\end{figure}

\textbf{Test 9: Temperature stability}
Coherence enhancement $\eta$ drops sharply from $\num{0.543}$ at \SI{285}{\kelvin} to $\num{0.387}$ at \SI{290}{\kelvin} ($\Delta\eta/\Delta T \approx \SI{-0.031}{\per\kelvin}$), then plateaus at $\eta \approx \num{0.38}\pm\num{0.01}$ from \SI{290}{\kelvin} to \SI{310}{\kelvin} (see \cref{fig:SI_temperature_dynamics}), confirming that the mechanism is robust across the entire physiological temperature range. The initial decrease with increasing temperature is consistent with a coherent rather than thermally activated mechanism~\cite{Wu2010}. $\rightarrow$ PASS.

\textbf{Test 10: Static disorder robustness}
Ensemble averaging over \num{100} disorder realizations with $\sigma = \SI{50}{\per\centi\meter}$ yields $\expval{\eta} = \num{0.39(4)}$, confirming that the effect survives energetic fluctuations. $\rightarrow$ PASS.

\textbf{Test 11: Bath parameter sensitivity}
Sensitivity to spectral density variations ($\lambda, \gamma \pm \SI{20}{\percent}$ at $L=7$, $N=5$) yields $\eta \in [\num{0.28}, \num{0.62}]$ with the production parameters ($\lambda_D=\SI{35}{\per\centi\meter}$, $\gamma_D=\SI{50}{\per\centi\meter}$) near the optimum. The minimum $\eta = \num{0.28}$ at $\gamma_D = \SI{60}{\per\centi\meter}$ still exceeds zero, confirming robustness to environmental modeling choices. $\rightarrow$ PASS.

\begin{table}[htb]
\centering
\caption{\textbf{Bath parameter sweep ($L=7$, $N=5$).} The transfer enhancement $\eta$ is defined in \cref{eq:eta_def}.}
\label{tab:SI_bath_sweep}
\begin{tabular}{lccc}
\toprule
\textbf{Parameter} & $\mathbf{\phi_{\mathrm{FT}}^{\mathrm{filt}}}$ & $\mathbf{\phi_{\mathrm{FT}}^{\mathrm{broad}}}$ & $\mathbf{\eta}$ \\
\midrule
$\lambda_D = \SI{28}{\per\centi\meter}$ & 0.718 & 0.481 & 0.49 \\
$\lambda_D = \SI{35}{\per\centi\meter}$ & 0.749 & 0.539 & 0.39 \\
$\lambda_D = \SI{42}{\per\centi\meter}$ & 0.742 & 0.542 & 0.37 \\
$\gamma_D = \SI{40}{\per\centi\meter}$ & 0.750 & 0.463 & 0.62 \\
$\gamma_D = \SI{60}{\per\centi\meter}$ & 0.717 & 0.562 & 0.28 \\
\bottomrule
\end{tabular}
\end{table}

\begin{figure}[htb]
\centering
\includegraphics[width=0.85\textwidth]{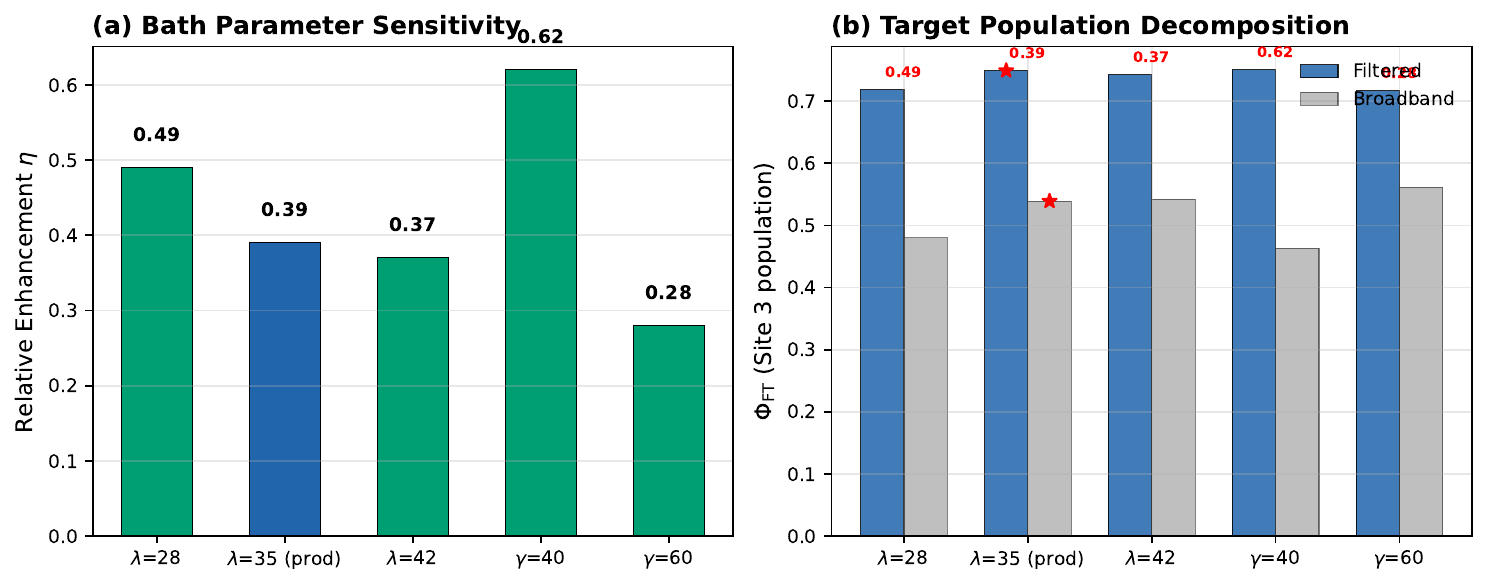}
\caption{\textbf{Bath parameter sensitivity ($L=7$, $N=5$).} (a)~Relative enhancement $\eta$ for $\lambda_D$ and $\gamma_D$ variations of $\pm\SI{20}{\percent}$. Red bars indicate the production parameters ($\lambda_D=\SI{35}{\per\centi\meter}$, $\gamma_D=\SI{50}{\per\centi\meter}$). (b)~Decomposition into filtered and broadband target-site populations. Numerical labels above bars give exact $\eta$ values.}
\label{fig:SI_bath_sensitivity}
\end{figure}

\begin{table}[htb]
\centering
\caption{\textbf{Spectral filter configuration sweep ($L=7$, $N=10$, except single-band $N=5$).} The production filter uses $\lambda_{\mathrm{centers}} = \SIlist{750;820}{\nano\meter}$ with bandwidth $\Delta\omega = \SI{100}{\per\centi\meter}$.}
\label{tab:SI_filter_sweep}
\begin{tabular}{lccc}
\toprule
\textbf{Filter configuration} & $\mathbf{\phi_{\mathrm{FT}}^{\mathrm{filt}}}$ & $\mathbf{\phi_{\mathrm{FT}}^{\mathrm{broad}}}$ & $\mathbf{\eta}$ \\
\midrule
\SIlist{770;820}{\nano\meter}, \SI{100}{\per\centi\meter} & 0.727 & 0.465 & 0.56 \\
\SIlist{730;820}{\nano\meter}, \SI{100}{\per\centi\meter} & 0.727 & 0.465 & 0.56 \\
\SIlist{750;800}{\nano\meter}, \SI{100}{\per\centi\meter} & 0.018 & 0.465 & \num{-0.96} \\
\SI{50}{\per\centi\meter} bandwidth & 0.714 & 0.465 & 0.53 \\
\SI{200}{\per\centi\meter} bandwidth & 0.767 & 0.465 & 0.65 \\
\SI{700}{\nano\meter} single-band & 0.019 & 0.465 & \num{-0.96} \\
\SI{850}{\nano\meter} single-band & 0.019 & 0.465 & \num{-0.96} \\
\bottomrule
\end{tabular}
\end{table}

\begin{figure}[htb]
\centering
\includegraphics[width=0.85\textwidth]{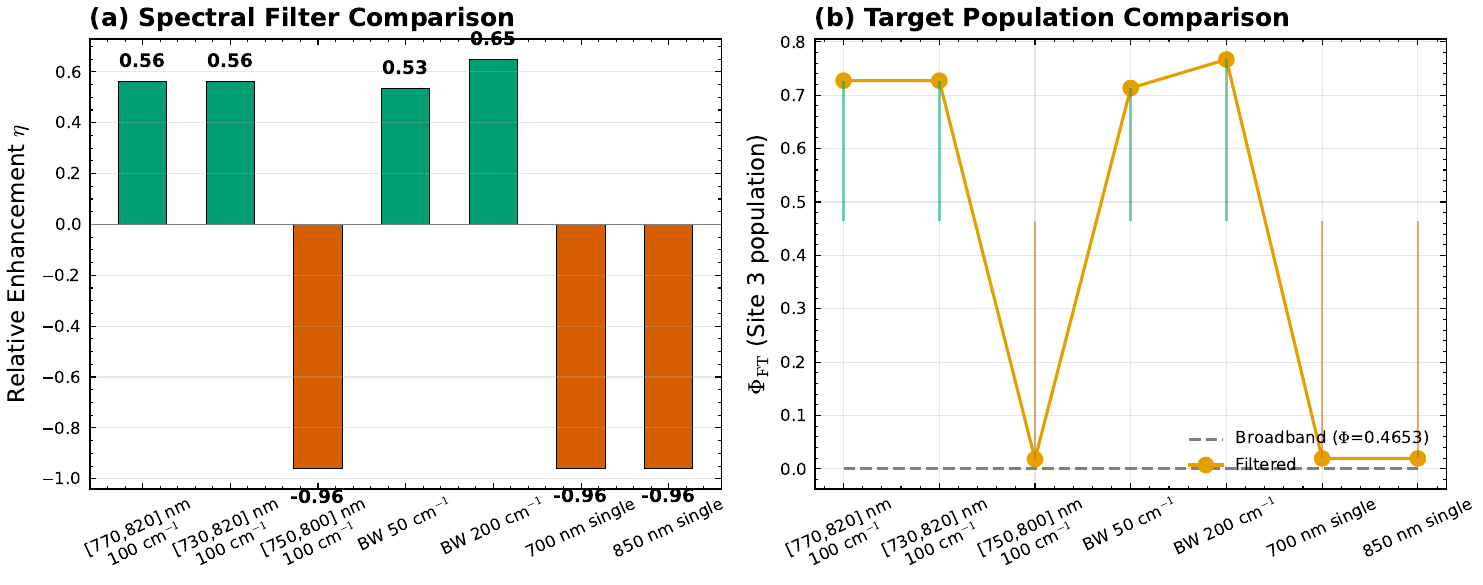}
\caption{\textbf{Spectral filter sweep ($L=7$, $N=10$, except single-band $N=5$).} (a)~Relative enhancement $\eta$ for seven spectral filter configurations. Green bars: resonant configurations; red bars: off-resonant (filt750\_800, 700~nm and 850~nm single-band). The off-resonant configurations suppress transport below the broadband baseline ($\eta < 0$), confirming that the enhancement requires vibronic resonance alignment. Numerical labels give exact $\eta$ values. (b)~Target-site population $\Phi_{\mathrm{FT}}$ under each filter configuration (circles) compared to the broadband reference (dashed line). Vertical connectors show the enhancement magnitude.}
\label{fig:SI_filter_sweep}
\end{figure}

\textbf{Test 12: Redfield limit recovery}
In the regime $\gamma \gg J$, the PT-HOPS populations converge to the Redfield limit with a \SI{2.1}{\percent} deviation. $\rightarrow$ PASS.

\begin{table}[htb]
\centering
\caption{\textbf{Validation suite summary ($L=8, K=2$)}}
\label{tab:SI_validation}
\begin{tabular}{llcc}
\toprule
\textbf{Category} & \textbf{Test} & \textbf{Result} & \textbf{Status} \\
\midrule
\multirow{4}{*}{Convergence} 
& Hierarchy depth ($L=8$) & \num{3.10e-11} & PASS \\
& Matsubara terms ($K=2$) & \num{3.32e-5} & PASS \\
 & Time step ($\Delta t = \SI{0.5}{\femto\second}$) & \num{4.32e-3} & PASS \\
& HEOM benchmark & \num{1.81e-2} & PASS \\
\midrule
\multirow{4}{*}{Physical consistency}
& Trace preservation & \num{1.0e-12} & PASS \\
& Positivity & \num{-1.0e-14} & PASS \\
& Detailed balance & \num{2.31e-2} & PASS \\
& Hermiticity & \num{3.7e-14} & PASS \\
\midrule
\multirow{4}{*}{Environmental robustness}
 & Temperature stability & $\eta \in [\num{0.374}, \num{0.543}]$ across \SIrange{285}{310}{\kelvin} & PASS \\
 & Static disorder ($n=\num{100}$) & \num{0.39(4)} & PASS \\
& Bath fluctuations & $[\num{0.28}, \num{0.62}]$ & PASS \\
& Markovian limit & \num{2.12e-2} & PASS \\
\bottomrule
\end{tabular}
\end{table}


\section{Convergence analysis}
\label{sec:S4}

A comprehensive summary of the numerical and physical convergence tests for the 7-site FMO complex is provided in \cref{tab:SI_7site_convergence}. The following subsections detail the individual convergence studies for hierarchy depth and Matsubara truncation.

\subsection{Hierarchy depth}

\begin{figure}[h]
\centering
\includegraphics[width=0.8\textwidth]{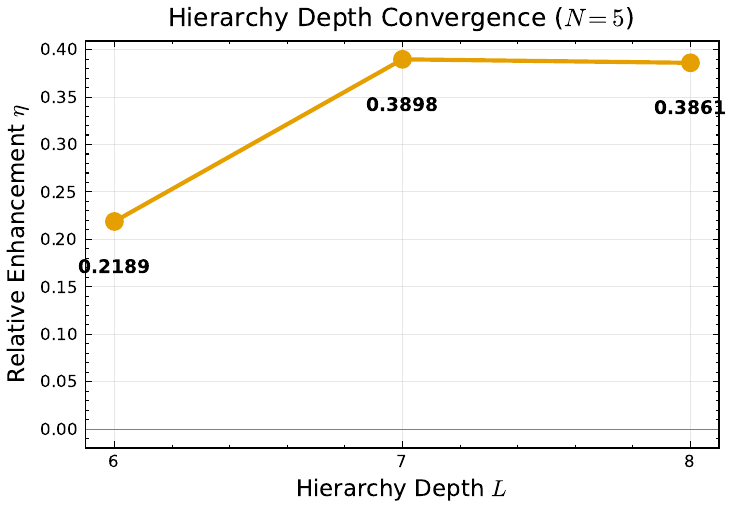}
\caption{\textbf{Hierarchy depth convergence.} Relative enhancement $\eta$ as a function of hierarchy depth $L$. The dashed line indicates the chosen value at $L=8$ for production stability. Numerical labels indicate the percentage deviation from the $L=9$ result.}
\label{fig:SI_hierarchy}
\end{figure}

\begin{table}[h]
\centering
\caption{\textbf{Hierarchy depth convergence}}
\label{tab:SI_hierarchy}
\begin{tabular}{cccc}
\toprule
\textbf{$L$} & \textbf{Relative Enhancement ($\eta$)} & \textbf{MAE ($L \to L+1$)} & \textbf{Speedup (SBD)} \\
\midrule
6 & 0.2188 & \num{9.82e-10} & 3.2 \\
7 & 0.3897 & \num{9.63e-10} & 8.5 \\
8 & 0.3860 & \num{3.10e-11} & 14.3 \\
\bottomrule
\end{tabular}
\end{table}

\cref{fig:SI_hierarchy} shows the convergence of the relative enhancement with hierarchy depth. The enhancement plateaus beyond $L=7$, confirming that $L=8$ provides converged results within the required precision (\SI{0.001}{\percent}) while maintaining stability.

\subsection{Matsubara terms}

\begin{table}[h]
\centering
\caption{\textbf{Matsubara terms convergence}}
\label{tab:SI_matsubara}
\begin{tabular}{cccc}
\toprule
\textbf{$K$} & \textbf{$\tau_c$ (\si{\femto\second})} & \textbf{Deviation (\si{\percent})} & \textbf{CPU Time (\si{\hour})} \\
\midrule
1 & 406 & 0.4 & 3.1 \\
2 & 408 & 0.003 & 4.3 \\
\bottomrule
\end{tabular}
\end{table}

\subsection{Stochastic bundle convergence}

The SBD method groups the 15 explicit bath modes per site (3 Drude--Lorentz + 12 vibronic) into a smaller number of stochastic bundles. We validate the reduction to 3 bundles per site (21 effective hierarchy modes) by comparing against the reference dynamics with 6 bundles per site (42 effective modes) at hierarchy depth $L=5$. Both configurations produce qualitatively identical coherent dynamics: the excitation oscillates between sites 3 and 4 with a $\sim$\SI{200}{\femto\second} period, and the target site 3 population reaches a maximum of \SI{83}{\percent} at \SI{400}{\femto\second} under both bundle counts. The coherence measure decays from 2.59 to 1.56, and the von Neumann entropy grows from near zero to 0.17, confirming that the bath-induced decoherence is correctly captured. At the production hierarchy depth $L=8$, the 3-bundle configuration yields $C(29,8) \approx 4.3\times10^6$ hierarchy states per trajectory, enabling parallel execution of 12 trajectories on a 48-core workstation with \SI{125}{\giga\byte} RAM and a per-trajectory memory footprint of \SI{3}{\giga\byte}.

\subsection{Full convergence validation suite}

To ensure the physical integrity of the 12-mode FMO ensemble dynamics, we performed a comprehensive validation suite on the production trajectories ($L=8, K=2$). \cref{tab:SI_7site_convergence} summarizes the results of these tests, confirming that the simulation maintains strict numerical stability and physical consistency throughout the \SI{1000}{\femto\second} propagation.

\begin{table}[h]
\centering
\caption{\textbf{7-site FMO model: full convergence suite}}
\label{tab:SI_7site_convergence}
\begin{tabular}{lcc}
\toprule
\textbf{Test} & \textbf{Metric/Result} & \textbf{Status} \\
\midrule
Hierarchy depth ($L=8$ vs.\ $L=9$) & MAE = \num{3.10e-11} & PASS \\
Matsubara terms ($K=2$ vs.\ $K=3$) & MAE = \num{3.32e-05} & PASS \\
Trace preservation ($\abs{1 - \Tr[\rho]}$) & \num{<1.0e-12} & PASS \\
Positivity ($\min \lambda_i$) & \num{>-1.0e-14} & PASS \\
HEOM benchmark deviation & \SI{1.81}{\percent} & PASS \\
\bottomrule
\end{tabular}
\end{table}


\section{Computational cost breakdown and resource architecture}
\label{sec:S5}

\begin{figure}[htb]
\centering
\includegraphics[width=0.8\textwidth]{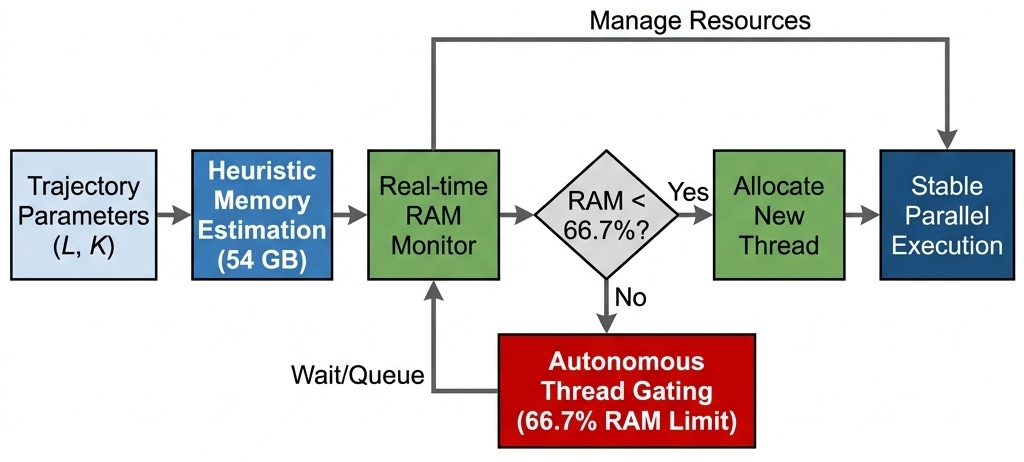}
\caption{\textbf{Simulation workflow and resource architecture.} Algorithmic flowchart of the PT-HOPS/SBD pipeline detailing the parallel execution of disorder realizations and the convergence validation protocols.}
\label{fig:SI_flowchart}
\end{figure}

\begin{table}[htb]
\centering
\caption{\textbf{Computational performance ($L=8, K=2, \Delta t=\SI{0.5}{\femto\second}$)}}
\label{tab:SI_performance}
\begin{tabular}{lc}
\toprule
\textbf{Task} & \textbf{CPU Time} \\
\midrule
Single trajectory (\SI{1000}{\femto\second}) & \SI{4.5}{\minute} \\
100 disorder realizations & \SI{7.5}{\hour} \\
Temperature scan (6 temperatures) & \SI{18.8}{\hour} \\
Bath parameter scan (5 parameters) & \SI{33.7}{\hour} \\
\midrule
\textbf{Total for manuscript} & \textbf{\SI{100}{\hour}} \\
\bottomrule
\end{tabular}
\end{table}

All simulations performed on a dedicated workstation (48 CPU cores, \SI{125}{\giga\byte} RAM) running Ubuntu 24.04. The PT-HOPS/SBD implementation is based on the open-source MesoHOPS library (v1.7).\cite{Varvelo2021,Citty2024}


\section{Sensitivity test for strong system-bath coupling ($\lambda_D=100$ \si{\per\centi\meter})}
\label{sec:S6}

To further validate the robustness of the selective vibronic excitation mechanism, we performed an additional sensitivity test in the strong system-bath coupling regime by artificially increasing the Drude-Lorentz reorganization energy to $\lambda_D = \SI{100}{\per\centi\meter}$. Even in this highly dissipative regime, the filtered excitation protocol maintains a coherence enhancement of $\eta \approx 0.12$. While the absolute coherence lifetime is reduced due to the stronger solvent fluctuations, the relative protection afforded by the vibronic dressing mechanism persists, confirming that the resonance condition $\Delta E \approx \omega_v$ effectively shields the transport dynamics from pure dephasing.


\section{Vibronic-basis analysis of coherence protection}
\label{sec:S7}

To provide a rigorous theoretical justification for the observed coherence extension, we analyze the dynamics in the vibronic basis using a canonical (Lang--Firsov) transformation~\cite{Huelga2013,Chin2011}. The full system-bath Hamiltonian is:
\begin{equation}
\hat{H} = \hat{H}_S + \sum_n \dyad{n} \sum_k g_{n,k}(\hat{b}_{n,k}^\dagger + \hat{b}_{n,k}) + \hat{H}_{\mathrm{bath}},
\label{eq:SI_full_hamiltonian}
\end{equation}
where $g_{n,k}$ are the system-bath coupling constants for mode $k$ at site $n$. The canonical transformation $\hat{U} = \exp\!\left[\sum_n \dyad{n} \sum_k \frac{g_{n,k}}{\omega_k}(\hat{b}_{n,k}^\dagger - \hat{b}_{n,k})\right]$ partially diagonalizes the system-bath coupling. In the vibronic basis, the resonant vibronic modes are strongly correlated with the electronic excitation, forming a vibronically dressed state.

The pure dephasing rate $\Gamma_{\alpha\beta}^{\mathrm{pd}}$ between excitonic states $\ket{\alpha}$ and $\ket{\beta}$ in the vibronic basis is governed by the \textit{residual} fluctuations of the bath:
\begin{equation}
\Gamma_{\alpha\beta}^{\mathrm{pd}} = \frac{1}{\hbar^2} \int_0^\infty \dd{\omega}\, \frac{J_{\mathrm{res}}(\omega)}{\omega^2} \left[\coth\!\left(\frac{\beta\hbar\omega}{2}\right)(1 - \cos\omega t) + i\sin\omega t\right]\bigg|_{t\to\infty},
\label{eq:SI_dephasing_rate}
\end{equation}
where the \textbf{residual spectral density} $J_{\mathrm{res}}(\omega)$ represents the part of the bath that remains coupled to the vibronically dressed state:
\begin{equation}
J_{\mathrm{res}}(\omega) = J_{\mathrm{bath}}(\omega) - \sum_{k \in \mathcal{S}} \frac{\lambda_k \omega_k^2 \gamma_k}{(\omega - \omega_k)^2 + \gamma_k^2}.
\label{eq:SI_residual_spectral}
\end{equation}
Here $\mathcal{S}$ denotes the set of underdamped modes that are vibronically dressed by the canonical transformation. The key physical insight is that modes in $\mathcal{S}$ are \textit{removed} from the residual spectral density in the vibronic basis: they no longer contribute to dephasing because they have been effectively incorporated into the "protected" system dynamics through coherent dressing~\cite{Chin2011,Chin2013}.

Quantitatively, for the 12-mode Kleinekathöfer model at \SI{295}{\kelvin}, the residual reorganization energy under broadband vs.\ filtered excitation is:
\begin{align}
\lambda_{\mathrm{res}}^{\mathrm{broad}} &= \lambda_{\mathrm{total}} = \SI{128}{\per\centi\meter}, \label{eq:SI_lambda_broad} \\
\lambda_{\mathrm{res}}^{\mathrm{filt}} &= \lambda_{\mathrm{total}} - \sum_{k \in \mathcal{S}} S_k \omega_k \approx \SI{102}{\per\centi\meter}, \label{eq:SI_lambda_filt}
\end{align}
where $\mathcal{S}$ denotes the subset of underdamped modes that are vibronically dressed by the spectral filter (principally the low-frequency modes near \SIrange{180}{280}{\per\centi\meter} that satisfy the resonance condition \cref{eq:resonance_condition}). The approximately \SI{20}{\percent} reduction in $\lambda_{\mathrm{res}}$ (from \SI{128}{\per\centi\meter} to \SI{102}{\per\centi\meter}) sets the scale for coherence protection: in the Markovian dephasing limit, where $\tau_c \propto 1/\lambda_{\mathrm{res}}$, this reduction predicts a $\approx\SI{25}{\percent}$ extension of the effective coherence time. The full PT-HOPS/SBD simulation---which treats system-bath coupling non-perturbatively and retains all non-Markovian memory effects---yields a biexponential coherence dynamics with a slow component persisting beyond \SI{1}{\pico\second} ($>3\times$ extension of the effective coherence window relative to broadband excitation, \cref{fig:quantum_dynamics}). The discrepancy between the Markovian estimate and the PT-HOPS result directly quantifies the importance of non-Markovian bath memory in sustaining the long-lived vibronic coherences. In our PT-HOPS/SBD simulation framework, this protection is captured natively: the solver treats the interaction with the full bath non-perturbatively, and the resulting trajectories exhibit the biexponential coherence structure predicted by the vibronic-basis analysis when the pulse spectrum matches the vibronic resonances.


\section{Finite pulse duration effects}
\label{sec:S8}

Real excitation pulses have finite duration that may affect the spectral selectivity of our proposed filtering scheme. Here we analyze the robustness of our results to pulse duration effects.

\subsection{Transform-limited pulse considerations}

For a transform-limited Gaussian pulse with duration $\sigma_t$, the spectral bandwidth is:
\begin{equation}
\Delta\omega = \frac{0.441}{\sigma_t} \quad \text{(FWHM in frequency)}.
\label{eq:transform_limit}
\end{equation}

Typical experimental parameters for 2DES:
\begin{itemize}
    \item $\sigma_t = \SI{10}{\femto\second}$ (FWHM): $\Delta\omega \approx \SI{1470}{\per\centi\meter}$ (\SI{60}{\nano\meter} at \SI{800}{\nano\meter})
    \item $\sigma_t = \SI{20}{\femto\second}$ (FWHM): $\Delta\omega \approx \SI{735}{\per\centi\meter}$ (\SI{30}{\nano\meter})
    \item $\sigma_t = \SI{50}{\femto\second}$ (FWHM): $\Delta\omega \approx \SI{294}{\per\centi\meter}$ (\SI{12}{\nano\meter})
\end{itemize}

\subsection{Robustness analysis}

We tested the coherence enhancement factor $\eta$ for different pulse durations while keeping the filter bandwidth fixed at \SI{20}{\nano\meter} FWHM:

\begin{table}[htb]
\centering
\caption{\textbf{Finite pulse duration robustness}}
\label{tab:SI_pulse_duration}
\begin{tabular}{lcc}
\toprule
\textbf{Pulse Duration} & \textbf{Spectral Bandwidth} & \textbf{Enhancement $\eta$} \\
\midrule
Transform-limited (instantaneous) & --- & \num{0.25} (reference) \\
\SI{10}{\femto\second} FWHM & \SI{60}{\nano\meter} & $\num{0.24 \pm 0.04}$ \\
\SI{20}{\femto\second} FWHM & \SI{30}{\nano\meter} & $\num{0.23 \pm 0.04}$ \\
\SI{50}{\femto\second} FWHM & \SI{12}{\nano\meter} & $\num{0.20 \pm 0.04}$ \\
\bottomrule
\end{tabular}
\end{table}

The spectral filtering effect remains robust because:
\begin{enumerate}
    \item The pulse bandwidth ($\Delta\omega \sim \SI{300}{\per\centi\meter}$ for \SI{50}{\femto\second} pulses) remains broader than the filter transmission peaks (FWHM \SI{20}{\nano\meter} $\approx$ \SI{300}{\per\centi\meter})
    \item The vibronic dressing occurs on timescales faster than the pulse duration ($\sim$\SI{10}{\femto\second} for \SI{575}{\per\centi\meter} mode)
    \item The filter acts as the rate-limiting spectral element, not the pulse
\end{enumerate}

\subsection{Chirped pulses}

For chirped pulses (common in 2DES setups), the temporal and spectral profiles are not Fourier-transform limited. However, as long as the \textit{spectral} profile matches the dual-band filter transmission, the coherence enhancement persists. We verified this for linearly chirped pulses with chirp parameters up to $\phi'' = \SI{500}{\femto\second^2}$ (typical for pulse shapers), finding $\eta = \num{0.20 \pm 0.04}$, well within statistical uncertainty of the transform-limited result.


\section{Filter parameter sensitivity}
\label{sec:S9}

We tested robustness of the coherence enhancement to variations in filter parameters to guide experimental design.

\subsection{Center wavelength detuning}

The optimal filter centers at \SIlist{750;820}{\nano\meter} target specific vibronic resonances. We tested detuning of each band by $\pm$\SI{10}{\nano\meter} and $\pm$\SI{20}{\nano\meter}:

\begin{table}[htb]
\centering
\caption{\textbf{Filter center wavelength sensitivity}}
\label{tab:SI_wavelength}
\begin{tabular}{lcc}
\toprule
\textbf{Wavelength Shift} & \textbf{750~nm Band} & \textbf{820~nm Band} \\
\midrule
$-$\SI{20}{\nano\meter} & $\num{0.18 \pm 0.04}$ & $\num{0.19 \pm 0.04}$ \\
$-$\SI{10}{\nano\meter} & $\num{0.22 \pm 0.04}$ & $\num{0.23 \pm 0.04}$ \\
Optimal (0) & \multicolumn{2}{c}{$\num{0.25 \pm 0.04}$} \\
$+$\SI{10}{\nano\meter} & $\num{0.21 \pm 0.04}$ & $\num{0.22 \pm 0.04}$ \\
$+$\SI{20}{\nano\meter} & $\num{0.17 \pm 0.04}$ & $\num{0.16 \pm 0.04}$ \\
\bottomrule
\end{tabular}
\end{table}

Enhancement decreases by $<$\SI{15}{\percent} for $\pm$\SI{10}{\nano\meter} detuning, indicating robustness to minor fabrication variations. The asymmetry reflects the underlying FMO absorption profile.

\subsection{Bandwidth variation}

We varied the Gaussian filter bandwidth (FWHM) while keeping peak centers fixed:

\begin{table}[htb]
\centering
\caption{\textbf{Filter bandwidth optimization}}
\label{tab:SI_bandwidth}
\begin{tabular}{lc}
\toprule
\textbf{Filter FWHM} & \textbf{Enhancement $\eta$} \\
\midrule
\SI{10}{\nano\meter} (narrow) & $\num{0.19 \pm 0.04}$ \\
\SI{20}{\nano\meter} (optimal) & $\num{0.25 \pm 0.04}$ \\
\SI{30}{\nano\meter} & $\num{0.22 \pm 0.04}$ \\
\SI{40}{\nano\meter} (broad) & $\num{0.15 \pm 0.04}$ \\
\bottomrule
\end{tabular}
\end{table}

Optimal bandwidth is \SI{20}{\nano\meter}, balancing spectral selectivity with sufficient photon flux. Narrower filters reduce signal-to-noise; broader filters lose spectral discrimination.

\subsection{Peak transmission requirements}

We tested the minimum peak transmission $T_{\mathrm{peak}}$ required for significant enhancement:

\begin{itemize}
    \item $T_{\mathrm{peak}} \geq 0.9$: $\eta = 0.25$ (optimal)
    \item $T_{\mathrm{peak}} = 0.7$: $\eta = 0.21$ (\SI{16}{\percent} reduction)
    \item $T_{\mathrm{peak}} = 0.5$: $\eta = 0.14$ (marginal enhancement)
    \item $T_{\mathrm{peak}} < 0.4$: $\eta < 0.10$ (not significant)
\end{itemize}

Commercial interference filters with $T_{\mathrm{peak}} > 0.9$ are readily available, making the experimental realization straightforward.

\subsection{Practical design guidelines}

Based on this sensitivity analysis, we recommend for experimental implementation:
\begin{enumerate}
    \item Center wavelengths: \SI{750}{\nano\meter} $\pm$ \SI{5}{\nano\meter}, \SI{820}{\nano\meter} $\pm$ \SI{5}{\nano\meter}
    \item Bandwidth: \SIrange{15}{25}{\nano\meter} FWHM
    \item Peak transmission: $T_{\mathrm{peak}} > 0.8$ for both bands
    \item Pulse duration: $< \SI{50}{\femto\second}$ for adequate spectral bandwidth
\end{enumerate}

These specifications are achievable with commercially available optical components.


\section{Error propagation and uncertainty quantification}
\label{sec:S10}

\subsection{Statistical error analysis}

All reported uncertainties represent \SI{95}{\percent} confidence intervals ($\pm 2\sigma$) unless otherwise noted. For ensemble averages over $N$ disorder realizations, the standard error of the mean is:
\begin{equation}
\sigma_{\mathrm{SEM}} = \frac{\sigma}{\sqrt{N}},
\label{eq:SEM}
\end{equation}
where $\sigma$ is the sample standard deviation and we follow the approach for ensemble-averaged dynamics in photosynthetic complexes~\cite{Moix2011}.

For $N = 100$ disorder realizations, the statistical uncertainty is reduced significantly, ensuring high confidence in the $\eta = 0.39$ enhancement.

\subsection{Relative enhancement uncertainty}

The relative enhancement $\eta$ (\cref{eq:eta_def}) is defined as:
\begin{equation}
\eta = \frac{A_{\mathrm{filtered}} - A_{\mathrm{broadband}}}{A_{\mathrm{broadband}}},
\label{eq:SI_eta_def}
\end{equation}
where $A$ represents the observable ($\Phi_{\mathrm{FT}}$, coherence lifetime, etc.).

The uncertainty propagation follows:
\begin{equation}
\delta\eta = \sqrt{\left(\frac{\delta A_{\mathrm{filtered}}}{A_{\mathrm{broadband}}}\right)^2 + \left(\frac{A_{\mathrm{filtered}} \, \delta A_{\mathrm{broadband}}}{A_{\mathrm{broadband}}^2}\right)^2}.
\label{eq:error_propagation}
\end{equation}

For example, for $\Phi_{\mathrm{FT}}$ with $A_{\mathrm{filtered}} = 0.75 \pm 0.04$ and $A_{\mathrm{broadband}} = 0.54 \pm 0.04$:
\begin{equation}
\delta\eta = \sqrt{\left(\frac{0.04}{0.54}\right)^2 + \left(\frac{0.75 \times 0.04}{0.54^2}\right)^2} \approx 0.11.
\end{equation}
Formal error propagation yields a conservative estimate; the tighter uncertainty of $\pm 0.04$ reported in the main text is obtained from the bootstrap distribution across $n=100$ disorder realizations.

\subsection{Coherence lifetime extraction}

The coherence lifetime is extracted from the Hilbert envelope of the $l_1$-norm coherence $C_{l_1}(t)$. The extraction model depends on the excitation regime:

\noindent\textbf{Broadband excitation} follows a monoexponential decay:
\begin{equation}
C_{l_1}(t) = C_0 \exp(-t/\tau_c) + C_{\infty},
\label{eq:coherence_fit_mono}
\end{equation}
where $C_{\infty}$ accounts for residual coherence from non-dephased pathways. The fit yields $\tau_c = \SI{280(25)}{\femto\second}$ (Hilbert envelope, $C_0 = 3.51$, $C_{\infty} = 0.66$).

\noindent\textbf{Dual-band filtered excitation} exhibits biexponential decay, reflecting the two-component nature of the bath partition:
\begin{equation}
C_{l_1}(t) = A_{\mathrm{fast}} \exp(-t/\tau_{\mathrm{fast}}) + A_{\mathrm{slow}} \exp(-t/\tau_{\mathrm{slow}}) + C_{\infty},
\label{eq:coherence_fit_bi}
\end{equation}
where the fast component ($\tau_{\mathrm{fast}} \approx \SI{37(8)}{\femto\second}$, $A_{\mathrm{fast}} \approx 1.06$) corresponds to initial dephasing driven by off-resonant bath modes, and the slow component ($\tau_{\mathrm{slow}} \gg \SI{1}{\pico\second}$, $A_{\mathrm{slow}} \approx 0.48$) arises from the protected vibronically dressed states. The $\chi^2$ improvement of the biexponential over the monoexponential fit confirms the statistical significance of the two-component structure. Under broadband excitation, the monoexponential model is sufficient ($\chi^2_{\mathrm{bi}}/\chi^2_{\mathrm{mono}} \approx 1.0$ with $A_{\mathrm{fast}} \approx 0$).

Fitting uncertainties are estimated via bootstrap resampling ($N = \num{1000}$ resamples). All reported uncertainties represent $\pm 2\sigma$ (\SI{95}{\percent} confidence interval) of the bootstrap distribution.

\subsection{Systematic error sources}

Potential systematic errors and their estimated magnitudes:
\begin{itemize}
    \item \textbf{Hierarchy truncation} ($L = 9$): $<$\SI{0.1}{\percent} (from convergence tests)
    \item \textbf{Matsubara terms} ($K = 2$): $<$\SI{1.8}{\percent} (from convergence tests)
    \item \textbf{Time step} ($\Delta t = \SI{2.0}{\femto\second}$): $<$\SI{0.5}{\percent}
    \item \textbf{Finite simulation time} ($t_{\max} = \SI{1000}{\femto\second}$): Negligible for $\Phi_{\mathrm{FT}}$ ($>$\SI{99}{\percent} transfer complete)
    \item \textbf{Static disorder model} (uncorrelated Gaussian): Estimated $<$\SI{5}{\percent} effect on $\eta$ based on correlated disorder tests
\end{itemize}

Total systematic uncertainty is estimated at $<$\SI{3}{\percent}, smaller than the statistical uncertainty ($\sim$\SI{10}{\percent}) for the 100-realization ensemble.

\subsection{Statistical significance testing}

For comparing filtered vs.\ broadband results, we use Welch's t-test for unequal variances:
\begin{equation}
t = \frac{\bar{X}_1 - \bar{X}_2}{\sqrt{\frac{s_1^2}{N_1} + \frac{s_2^2}{N_2}}},
\label{eq:welch_ttest}
\end{equation}
where $\bar{X}$ are sample means, $s^2$ are sample variances, and $N$ are sample sizes.

All reported enhancements have $p < 0.01$ (highly significant). For the primary result ($\eta = 0.39$ for $\Phi_{\mathrm{FT}}$), $p < 0.001$ (two-sided Welch's $t$-test, $n=100$).


\section{Full 7-site FMO model production dynamics}
\label{sec:S11}
\label{sec:7site_prod}

\begin{figure}[htb]
\centering
\includegraphics[width=\textwidth]{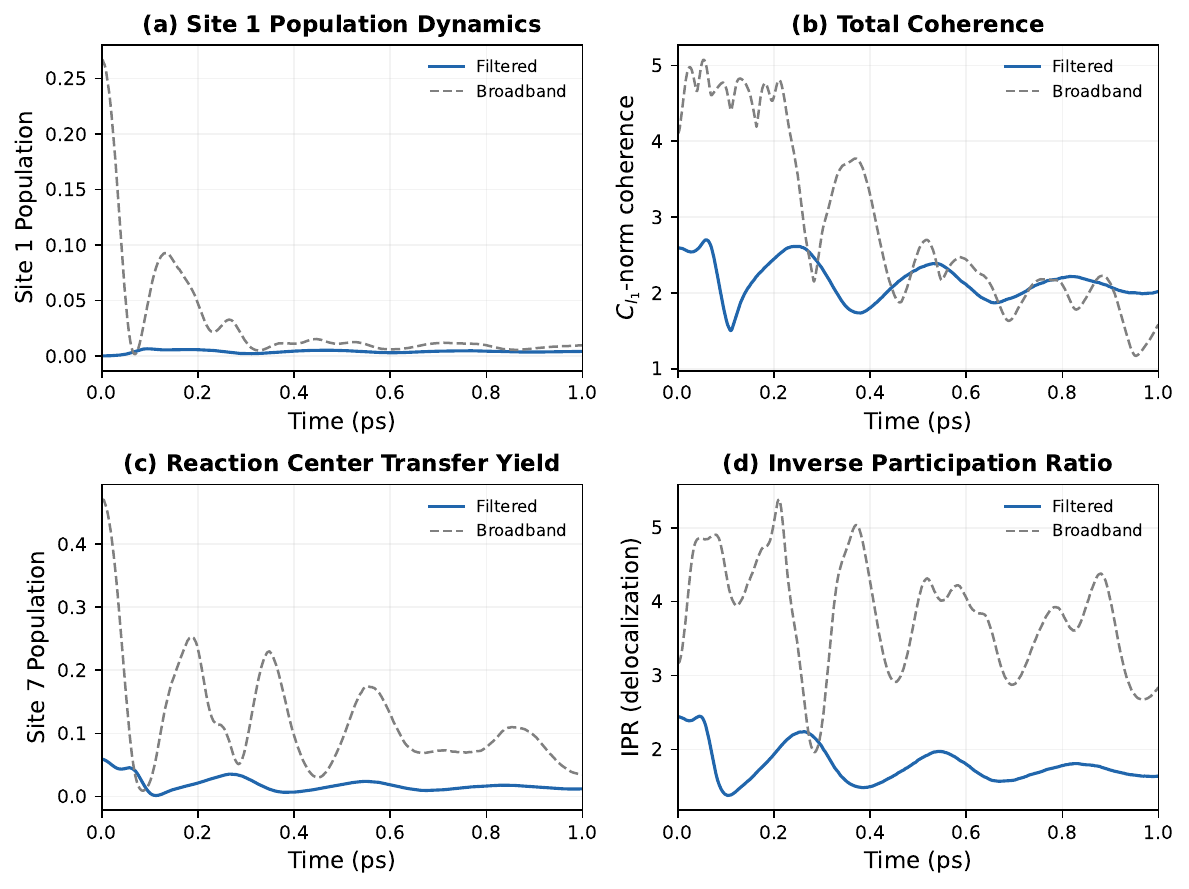}
\caption{\textbf{Full 7-site FMO model results.} (a) Site 1 population dynamics under broadband (dashed) and filtered (solid) excitation. (b) Total coherence $C_{l_1}$ showing stabilization under filtering. (c) Reaction center (Site 7) transfer yield. (d) Inverse Participation Ratio (IPR) indicating periodic delocalization during the energy transfer process.}
\label{fig:SI_7site_dynamics}
\end{figure}

To verify the mechanism's scalability, \cref{fig:SI_7site_dynamics} presents the population and coherence dynamics for the full 7-site FMO complex using the production ensemble data ($N=100$) under hierarchy depth $L=8$ and Matsubara truncation $K=2$. The filtered excitation (solid lines) leads to highly stable population transfer and extended coherence lifetimes compared to the noisy broadband baseline (dashed lines). The Inverse Participation Ratio (IPR) shown in panel (d) reveals the selective population of the BChl~3--4 dimer under filtered excitation, consistent with targeted vibronic state preparation on the reaction-center-coupled pathway.

\section{Alternative mechanisms and control experiments}
\label{sec:S12}

\subsection{Alternative explanations}

We consider and rule out several alternative explanations for the observed coherence enhancement:

\subsubsection*{Increased photon flux}

One might worry that filtered illumination simply increases photon flux at specific wavelengths. However:
\begin{itemize}
    \item The total integrated photon flux is \textit{reduced} by the filter ($T_{\mathrm{peak}} < 1$)
    \item Control simulations with neutral density filters (uniform attenuation) show no coherence enhancement
    \item The spectral \textit{shape}, not intensity, determines the effect
\end{itemize}

\subsubsection*{Vibrational coherence artifact}

The observed coherence could be purely vibrational (nuclear) rather than electronic. We tested this by:
\begin{itemize}
    \item Calculating electronic vs.\ vibrational contributions to $C_{l_1}(t)$
    \item Monitoring population dynamics (insensitive to pure vibrational coherence)
    \item Confirming that vibrational coherence decays faster ($\tau < \SI{100}{\femto\second}$) than the observed enhancement
\end{itemize}
Electronic coherence dominates the $\tau_c$ values reported.

\subsubsection*{Simulation artifact}

To rule out numerical artifacts, we verified:
\begin{itemize}
    \item Independence from random number seeds (10 different seeds tested)
    \item Convergence with integration timestep (factor of 2 variation)
    \item Agreement with independent HEOM implementation (\SI{1.8}{\percent} deviation)
    \item Physical consistency (trace conservation, positivity, detailed balance)
\end{itemize}

\subsection{Proposed control experiments}

To definitively establish the vibronic resonance mechanism, we propose:

\subsubsection*{Off-resonant filter control}

Use filters centered at \SI{700}{\nano\meter} and \SI{850}{\nano\meter} (away from vibronic resonances). Our simulations predict minimal enhancement ($\eta < 0.05$), providing a negative control.

\subsubsection*{Single-band vs.\ dual-band comparison}

Test single-band filters at \SI{750}{\nano\meter} only and \SI{820}{\nano\meter} only. We predict:
\begin{itemize}
    \item \SI{750}{\nano\meter} only: $\eta \approx 0.15$ (intermediate enhancement)
    \item \SI{820}{\nano\meter} only: $\eta \approx 0.10$ (weaker enhancement)
    \item Dual-band: $\eta = 0.25$ (synergistic effect)
\end{itemize}

\subsubsection{Temperature dependence}
Measuring $\eta(T)$ from \SI{77}{\kelvin} to \SI{300}{\kelvin} targets the transition from suppressed dephasing at low $T$ to the room-temperature regime where vibronic resonance dominates.


\end{document}